%% file: HFP.tex
\documentclass[
 aps, pra,
 amsmath,amssymb,
 11pt,
 final,
tightenlines,
 twoside,
 twocolumn,
 nofloats,
nofootinbib,
 superscriptaddress,
showkeys,
showkeywords,
 ]
{revtex4-2}

\usepackage[T1]{fontenc}
\usepackage[utf8x]{inputenc}
\usepackage[english]{babel}
\usepackage{graphicx}
\usepackage{dcolumn}
\usepackage{bm}
\usepackage{longtable}
\usepackage{appendix}

\input{maik.rty}

\setcitestyle{authoryear,round}
\input{sao_cmd_author.tex}

\begin{document}

\selectlanguage{english}



\title{High Frequency Peak Radio Sources from the AT20G Catalogue and Their Radio Spectra}

\author{\firstname{E.}~\surname{Majorova}}
 \affiliation{\saoname}

\author{\firstname{O.}~\surname{Zhelenkova}}
 \affiliation{\saoname}
  \email{zhe@sao.ru}

\begin{abstract}
A sample of high-frequency peaker (HFP) candidates was formed from the AT20G catalog radio sources with spectral indices of the optically thick emission region $\alpha_{below}$ exceeding +0.5.
A study of the spectral properties of the sources in the sample, which included 269 radio sources, was performed.
The spectra of the sources were constructed and the spectral indices below $\alpha_{below}$ and above the peak $\alpha_{above}$, the peak frequency $\nu_{obs}$, the flux density at the peak frequency $S_{peak}$, and the peak half-width in the radio spectrum were determined.
Analysis of the spectra showed that the sample is fairly homogeneous and consists of HFPs with $\nu_{obs}>5$\,GHz. Most sources (67\%) do not have data at frequencies below 0.8\,GHz.
187 sources have ultra-inverted spectra ($\alpha_{below}>$+0.7), which is 3.2\% of all sources in the AT20G catalog and 70\% of radio sources in our sample. Optical identification of radio sources in the sample showed that 70\% of the hosts are quasars.
The sample consists of compact objects with radio luminosity at 20 GHz in the range of \mbox{$10^{23}$--$10^{30}$\,W/Hz}, angular sizes of emitting regions of radio sources are 0.002--0.25\,mas, projected linear sizes are from 0.2 to 30\,pc.
The dependence of the peak frequencies of radio sources on their angular sizes is in good agreement with that previously discovered for CSS and GPS sources.
\end{abstract}

\maketitle

\section{INTRODUCTION}

Radio sources with convex spectra, which include CSS (Compact Steep Spectrum), GPS (Gigahertz-Peaked Spectrum) and HFP (High Frequency Peaker) sources, are related by an evolutionary process and are considered to be predecessors of FRI and FRII radio galaxies~\citep{1974MNRAS.167P..31F}.
They differ in the frequency of the peak in the spectrum, compactness and radio luminosity.
The frequency of the spectral peak of CSS sources lies in the range of $\nu_{max}$<0.5\,GHz, for GPS -- 0.5<$\nu_{max}$<5\,GHz and HFP -- $\nu_{max}$>5\,GHz; the sizes of CSS are from 1 to 20\,kpc, for GPS/HFP -- less than 1\,kpc.

It is believed that HFP sources are the predecessors of GPS, which in turn evolve into CSS objects. The final stage of the development of the GPS and CSS radio source population are FRI/FRII radio galaxies.
Studying radio sources with a peak in the radio spectrum is important for understanding the evolution of radio galaxies and their active nuclei (AGN), as well as the influence of the interstellar medium (ISM) on the evolution of the radio source.

At the early stages of the evolution of the radio source, the  jet interacts with the inhomogeneous and fairly dense interstellar medium near the active galactic nucleus.
Later, the jets end up in a more rarefied intergalactic medium and, propagating in it, generate radio sources whose sizes in some cases reach several megaparsecs. In such a scenario, the size of the radio source indicates the stage of its evolution and age.

The peak in the continuum radio spectrum of GPS/HFP sources is usually associated with synchrotron self-absorption\footnote {Synchrotron self-absorption -- scattering of photons on synchrotron electrons.} (SSA)~ \citep{1990A&A...231..333F,2000MNRAS.319..445S} or with free-free absorption (FFA)~ \citep{2003MNRAS.346..327T,2003A&A...401..113V}.

A spectrum with a peak at frequencies from 0.5 to tens of GHz is a characteristic feature of ``young'', compact radio sources.
They are distinguished by high radio luminosity (up to $10^{29}$ W/Hz), small size (less than 1\, kpc), weak variability in the radio range ($\le10\%$ per year) and a low degree of polarization~\citep{1991ApJ...380...66O, 1982ApJ...255...39R, 1998PASP..110..493O, 2005A&A...432...31T}.
It is believed that the youngest radio sources may exhibit variability in the optically thick region of the emission due to its evolution~\citep{2008A&A...487..885O} or changes in the properties of the absorbing medium through which the emission passes~\citep{2003AJ....126..723T}.

Young radio sources are rare due to the short lifespan of this stage of the radio source's development. They are ideal objects for studying and understanding the processes occurring at the early stages of the radio source's evolution.

According to the ``youth'' scenario described in~\cite{1998PASP..110..493O} and also in~\cite{1995A&A...302..317F, 1996ApJ...460..634R,2000MNRAS.319..445S}, there is a relationship between the age of the source and the frequency of the peak in the spectrum, so that the youngest radio sources should be sought among those whose continuum radio spectrum has a maximum at frequencies above a few gigahertz.
The most suitable candidates for the role of young radio sources are HFPs.

\section{AT20G survey}

The AT20G survey is a blind survey of the southern sky at 20\, GHz, which was carried out with the ATCA (Telescope Compact Array) from 2004 to 2008.
It covers almost the entire southern sky (6.1 sr), except for the |b|<$1.5^{\circ}$ band. To obtain reliable information on the spectral indices of the sources, observations were carried out almost simultaneously at 4.8 and 8.6\, GHz. Observations of 5890 sources were carried out in both full intensity and polarization.
Detailed information on the AT20G studies is given in~\cite{2010MNRAS.405.1560M, 2011MNRAS.412..318M, 2011ExA....32..147H}.

The AT20G catalogue, compiled from the survey, after excluding galactic clouds, planetary nebulae and HII regions, includes 5808 sources and gives the most complete picture of the high-frequency radio source population\footnote {On the Internet, the AT20G catalogue is available through the Vizier database (http://vizier.u-strasbg.fr)}.
It includes an order of magnitude more sources than previous catalogues at high frequencies and consists mainly of radio galaxies~\citep{2010MNRAS.405.1560M, 2011MNRAS.412..318M}.
The catalogue has a spectral flux density limit\footnote {Hereinafter referred to as the flux density.} of 40 mJy at 20 GHz with a catalogue completeness of 91\%.

The AT20G catalog contains the most complete sample of radio sources with predominantly high-frequency radio emission and provides rich material for searching for "young" sources. In this work, we used the AT20G catalog to create a sample of GPS/HFP candidates.

\section{Selection of GPS/HFP Candidates}

A characteristic feature of GPS/HFP sources is a peak in the radio spectrum. Such a spectrum corresponds to the theoretical spectrum of synchrotron radiation with self-absorption at low frequencies for a compact radio source with a uniform distribution of matter and magnetic field.

In selecting GPS/HFP candidates, we used the ``canonical'' spectrum criteria from~\cite{1991ApJ...380...66O, 1997A&A...321..105D, 1981ARA&A..19..373K}, according to which the spectral indices below and above the $\nu_{obs}$ peak in the spectrum are $\alpha_{below}$=+0.5 and $\alpha_{above}$=-0.7, respectively. $\alpha_{below}$ and $\alpha_{above}$ characterize optically thick and optically thin regions of emission.
Another parameter considered as a criterion of the "canonical" spectrum is the width of the spectrum peak at half the emission power FWHM (full width at half maximum).
In \cite{1991ApJ...380...66O, 1997A&A...321..105D, 2004A&A...424...91E} FWHM was taken equal to $\sim 1.2$ frequency decades.

GPS sources are radio sources with a peak in the radio spectrum in the range of 0.5\,GHz<$\nu_{obs}$<5\,GHz.
Objects with a peak in the spectrum at $\nu_{obs}$>5\,GHz are classified as HFP~\citep{2000A&A...363..887D, 2013AstBu..68..262M}\footnote{In~\cite{2019AstBu..74..348S} objects with a peak frequency at $\nu_{int}$>5\,GHz, where $\nu_{int}$ is the peak frequency in the source's reference frame, are considered HFP.}.

In selecting radio sources as GPS/HFP candidates, we used the following main criterion $\alpha_{below}\geq$+0.5. For this purpose, we calculated the two-frequency spectral indices 
$\alpha^{1}_{20}$, $\alpha^{5}_{20}$, $\alpha^{5}_{9}$, and $\alpha^{9}_{20}$ for each AT20G source\footnote{$\alpha^{1}_{20}$ is the spectral index between 1 and 20 GHz, $\alpha^{5}_{20}$ is the spectral index between 4.8 and 20 GHz, $\alpha^{5}_{9}$ is the spectral index between 4.8 and 8.6 GHz, and $\alpha^{9}_{20}$ is the spectral index between 8.6 and 20 GHz.}.
Then, for sources with $\alpha^{1}_{20}$, $\alpha^{5}_{20}$, $\alpha^{5}_{9}$, $\alpha^{9}_{20}$ $\geq +0.5$, their spectra were constructed using the web interface of the CATS database\footnote{https://www.sao.ru/cats}~\citep{1993BSAO...36..132V, 2005BSAO...58..118V} and the information contained therein.

Of the approximately 1000 spectra of sources viewed, 620 sources were selected for further analysis. The remaining radio sources were rejected because they did not meet the requirements already at the stage of visual inspection of these preliminary spectra.

Then, the spectra of the selected sources were analyzed using the spg program\footnote{The spg program is part of the FADPS~\citep {1997ASPC..125...46V} package of the RATAN-600 radio telescope radio astronomy data processing system.}.
The spectra of these sources were constructed by approximation with linear or parabolic functions on a logarithmic scale. Spectral indices $\alpha_{below}$ were determined by linear approximation of the low-frequency region of the spectra (below the peak).

The final sample included 269 radio sources with a peak in the spectrum or with inverted spectra for which $\alpha_{below}\geq +0.5$. 

The radio sources included in the sample are presented in the table~\ref{tab:Sp.45+} in the appendix~\ref{Tab}. The description of the columns is also given there.
Examples of spectra with approximating curves are shown in the figures in the appendix~\ref{Fig}.

\section{Estimates of spectral parameters}

For the sample of GPS/HFP candidates we determined the following spectral parameters: the frequency of the peak in the spectrum in the observer's rest frame $\nu_{obs}$, the spectral indices $\alpha_{below}$ and $\alpha_{above}$ below and above the peak, respectively, the flux density at the peak frequency $S_{peak}$, the half-width of the peak FWHM, expressed in decades of frequency\footnote
{FWHM = log($\nu_{1}$)-log($\nu_{2}$), where $\nu_{1}$ and $\nu_{2}$ are the frequencies at which the flux densities are equal to 0.5$\cdotp S_{peak}$.}. If the source redshift $z$ was available, the frequency of the spectral peak in the source's rest frame $\nu_{int}$ was calculated.

To determine the spectral parameters, the software package for processing radio astronomical data FADPS \citep{1993BSAO...36..132V, 1997ASPC..125...46V} was used.

$\nu_{obs}$, FWHM, $S_{peak}$ were determined by approximating the source spectrum with a parabolic function on a logarithmic scale (log-parabolic function): \\
$log S_{\nu} = a(log \nu)^2 + b log \nu +c $, \\ where $S_{\nu}$ is the flux density at the frequency $\nu$; the coefficients a, b, c are determined by the least-squares method. 

Spectral indices $\alpha_{below}$ and $\alpha_{above}$ were determined by approximating high-frequency and low-frequency regions of the spectra with linear functions on a logarithmic scale.

For sources with significant variability, the approximation was performed using the envelope of the convex spectrum. For inverted spectra without a peak in the spectrum, a linear approximation on a logarithmic scale was used.

The value of $\alpha_{above}$ was not determined for all sources in the sample due to the lack of information on flux densities at frequencies above 20 GHz.
For the same reason, for a number of sources, the value of $\alpha_{above}$ was determined using two-frequency spectral indices $\alpha^{9}_{20}$ (9 and 20\,GHz) or $\alpha^{20}_{95}$ (20 and 95\,GHz).

The use of observational data obtained at different telescopes and observation epochs, in some cases, leads to difficulties in approximating the spectra. For these reasons, spectral parameters for some sources are given for two approximation options.

The results of estimating the parameters $\nu_{obs}$, $\alpha_{below}$, $\alpha_{above}$, FWHM, $S_{peak}$ and $\nu_{int}$ of the spectra of radio sources in our sample are given in the table~\ref{tab:Sp.45+} in the appendix~\ref{Tab}.

\begin{figure*}
\centerline{
\vbox{
\hbox{
\centerline{
\includegraphics[angle=0,width=0.33\textwidth,clip]{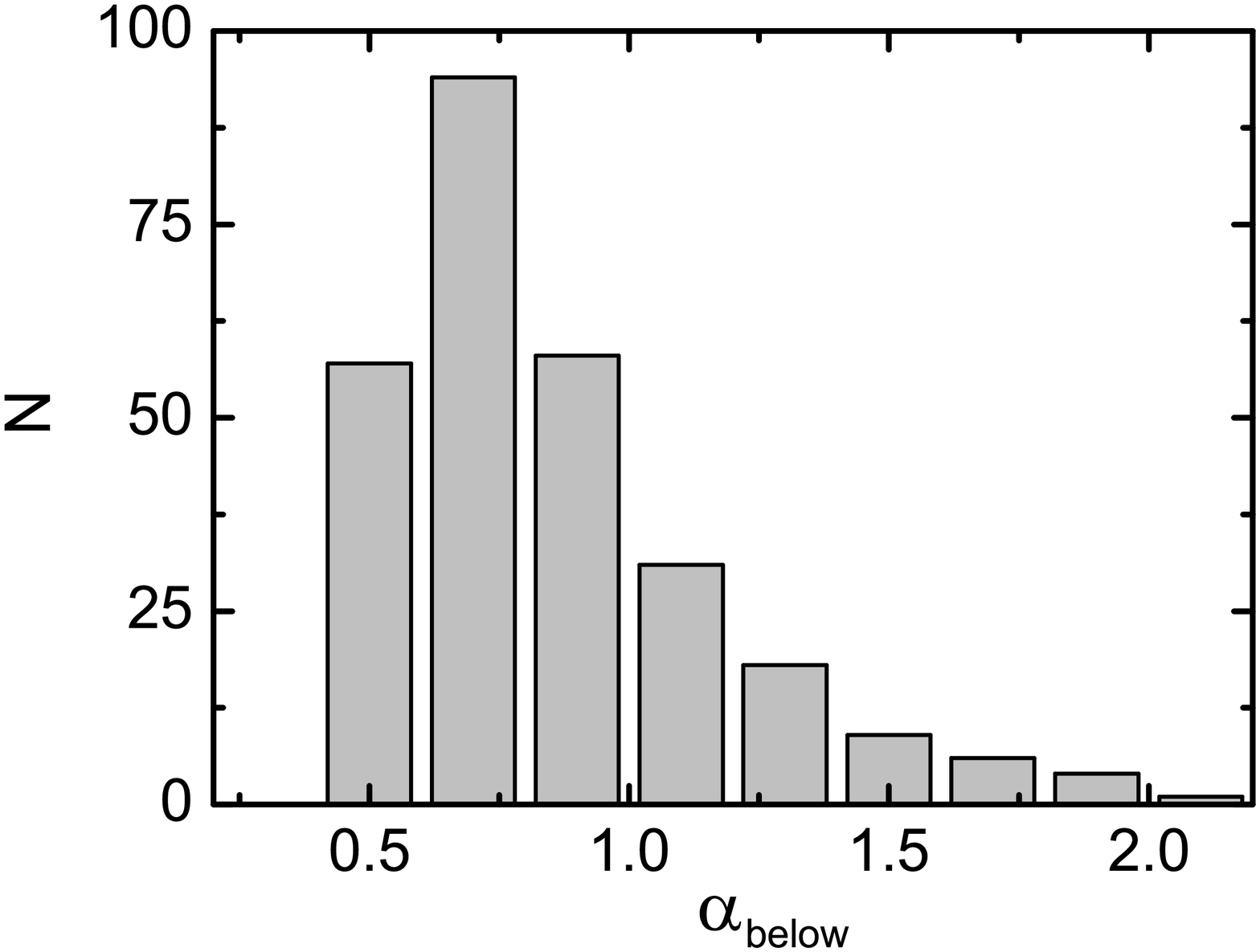}
\includegraphics[angle=0,width=0.33\textwidth,clip]{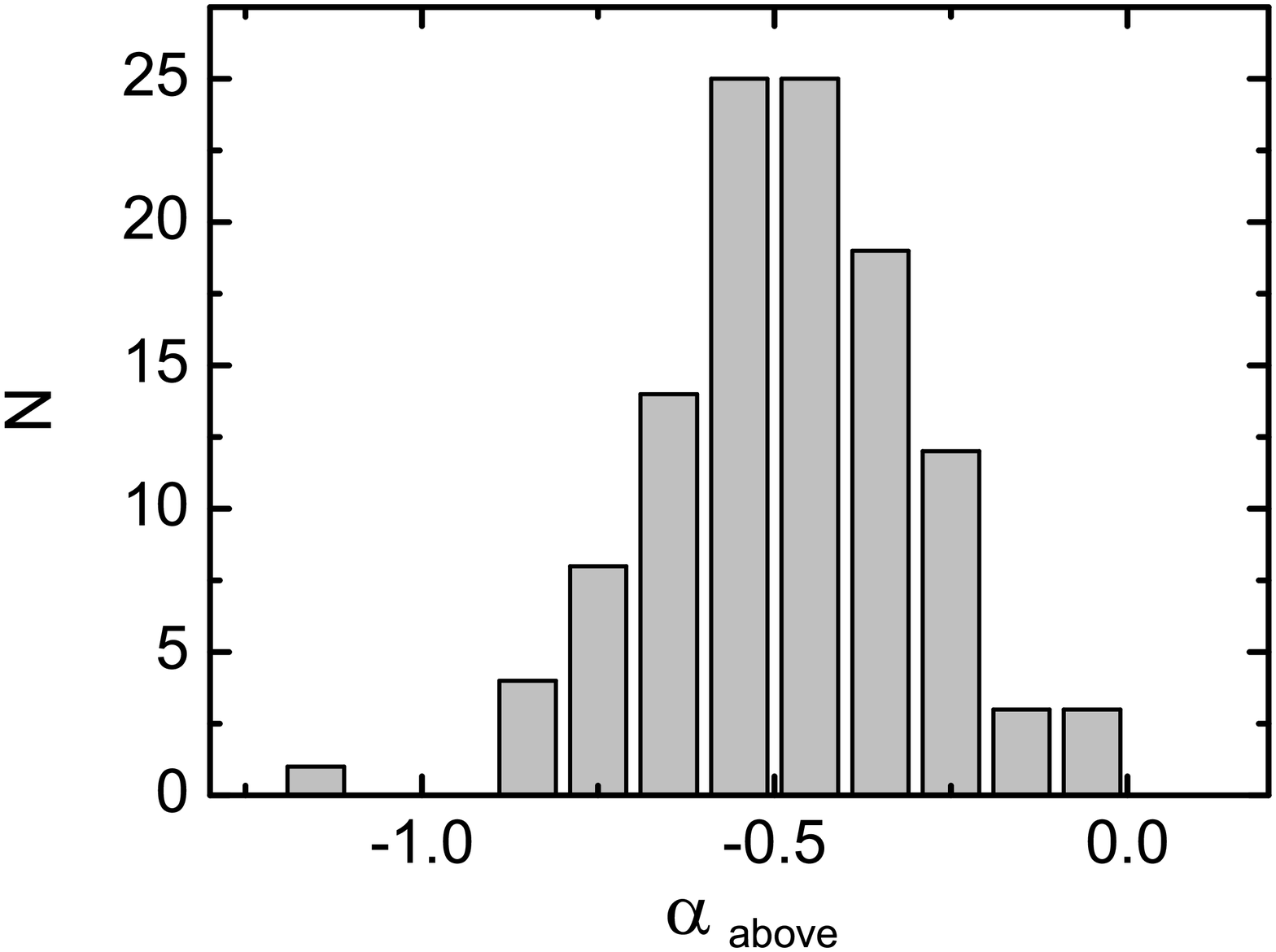}
\includegraphics[angle=0,width=0.33\textwidth,clip]{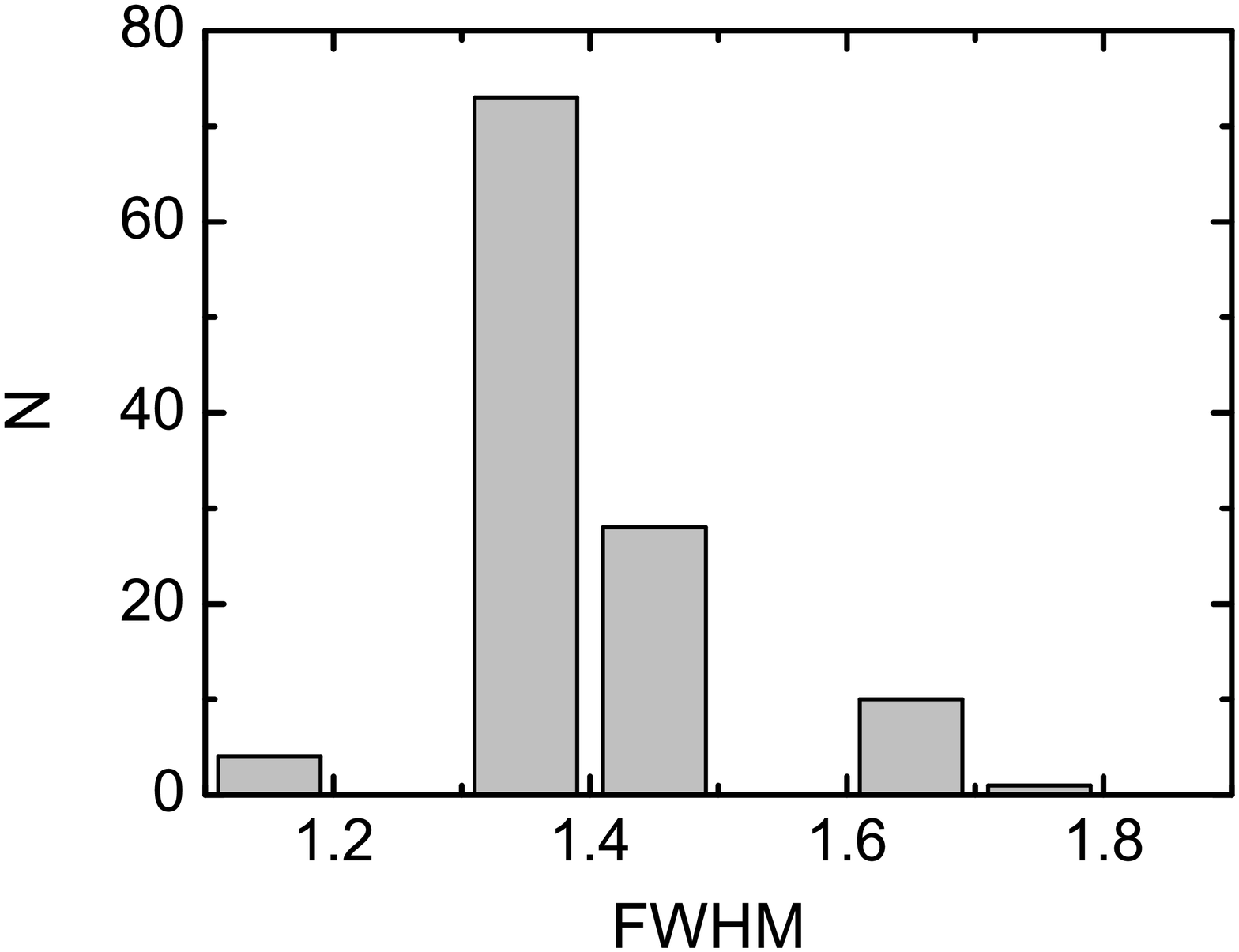}
}
}
\hbox{
\centerline{
\includegraphics[angle=0,width=0.33\textwidth,clip]{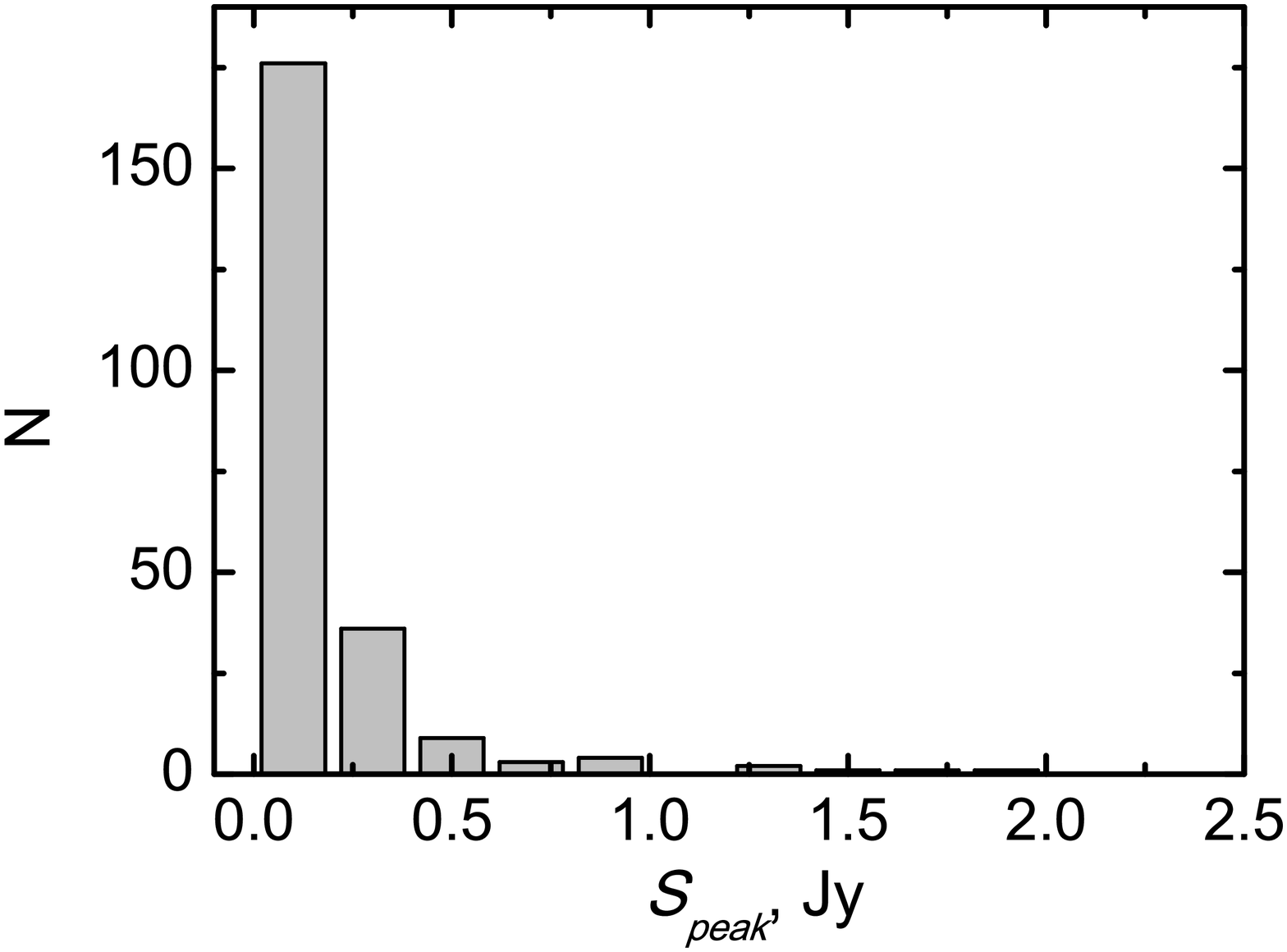}
\includegraphics[angle=0,width=0.33\textwidth,clip]{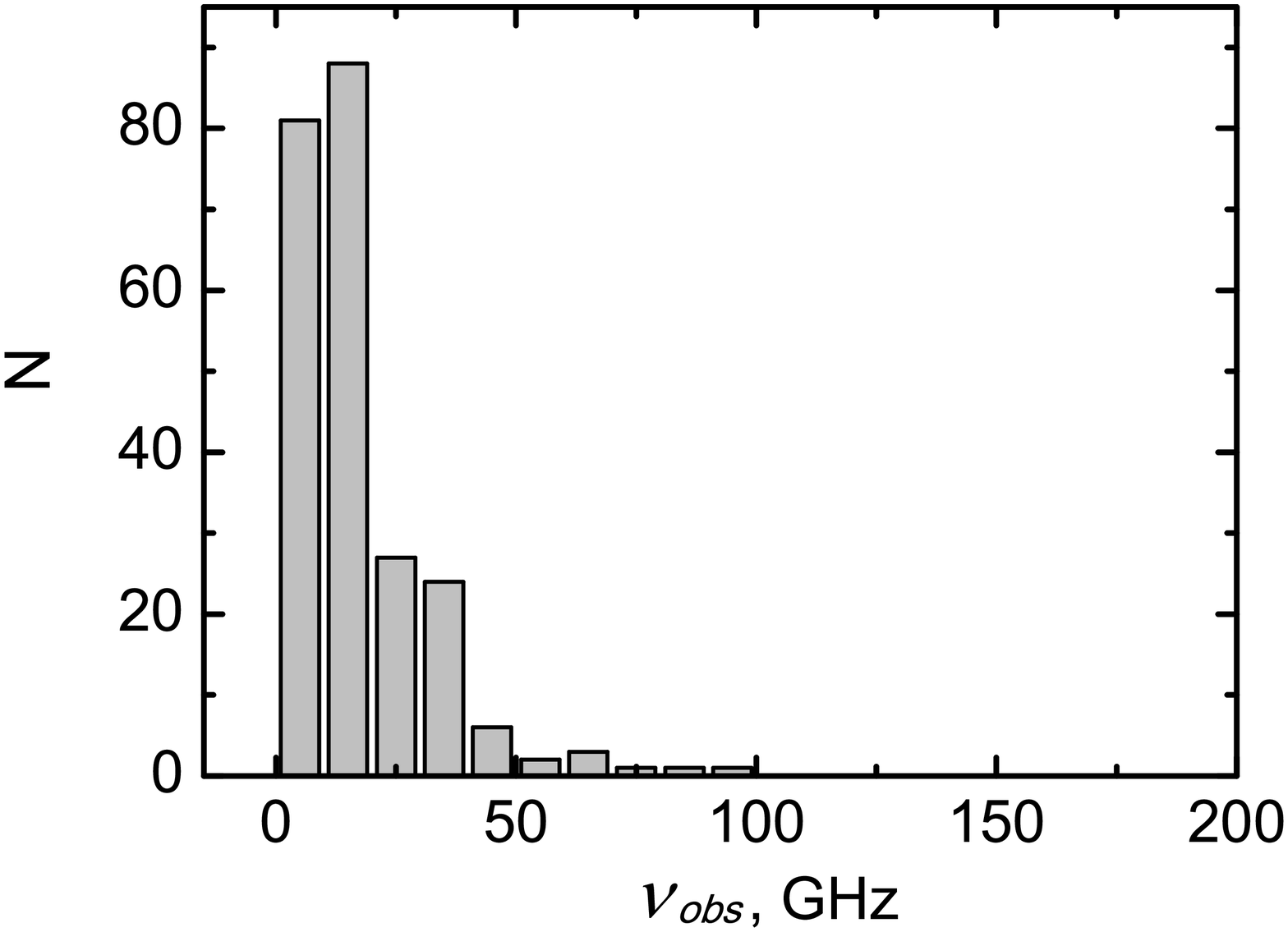}
\includegraphics[angle=0,width=0.33\textwidth,clip]{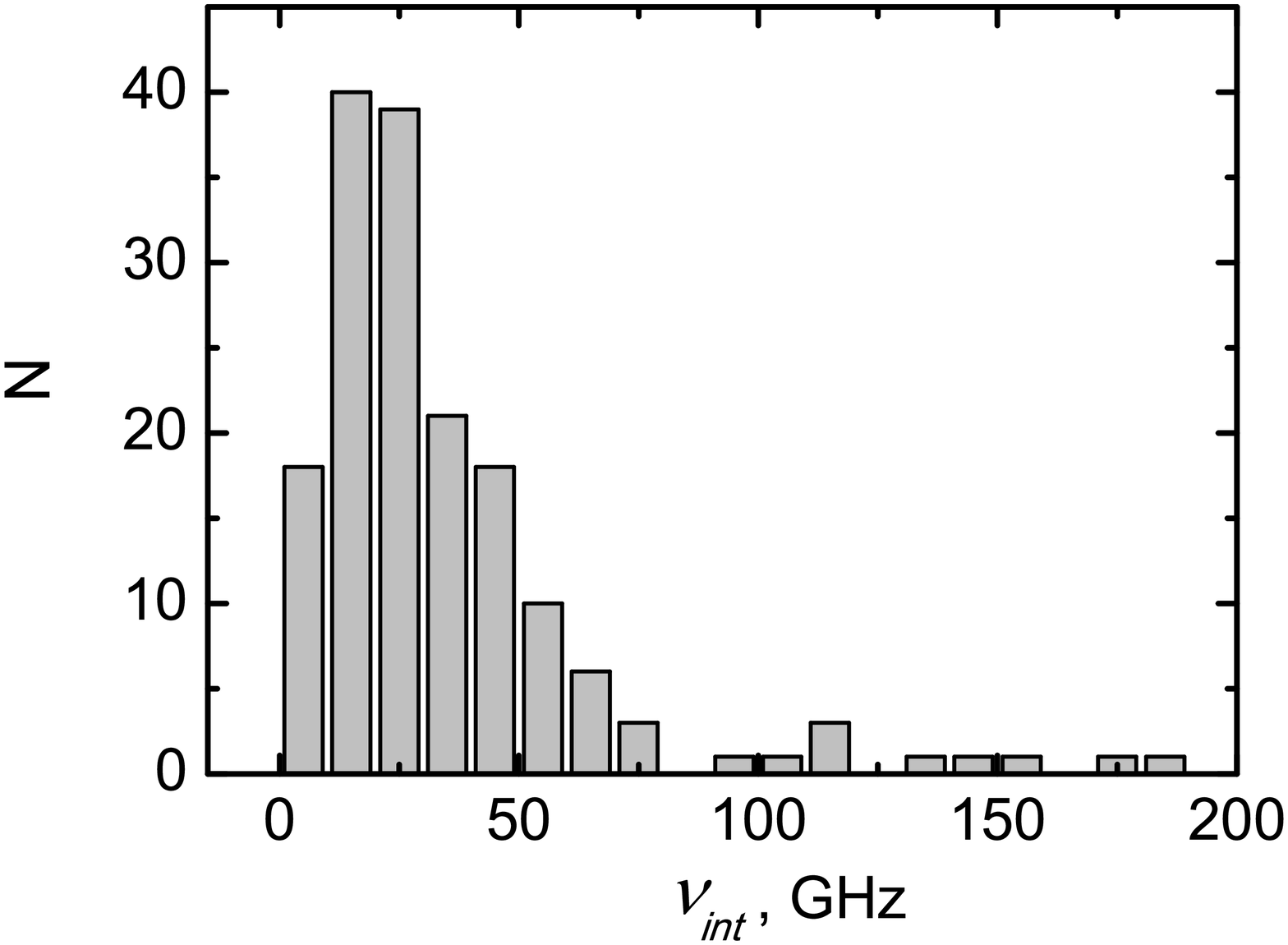}
}
}
}
}
\caption{
Histograms of the distributions of spectral indices $\alpha_{below}$, $\alpha_{above}$, FWHM, $S_{peak}$, $\nu_{obs}$, $\nu_{int}$ (from left to right and from top to bottom) of the radio sources of the sample.}
\label{fig1}
\end{figure*}
\section{ANALYSIS OF SPECTRA OF RADIO SOURCES}

Since we rely on heterogeneous data obtained with different instruments and at different observation epochs when constructing the source spectra, the spectra of some sources contain flux density values that bounce off the fitting curve, for example, J003207-154132 and J112120-172242 at 352 MHz in the WISH catalog or J011102-474911, J012744-345755, J025822-332705, J042119-672902 at 843 and 887 MHz in the SUMSS and RACS \citep{2021PASA...38...58H} catalogs.

By "bounces" we mean the flux density values, the inclusion of which in constructing the approximating spectrum curves significantly increases the root-mean-square error of the difference between the approximating curve and the flux densities. Such "bounces" may be associated with the variability of the source.
For example, for J134229-740728, the flux densities at 843 and 887 MHz differ by more than two times.
At the same time, the flux density at 887 MHz fits well into the linear spectrum constructed using the AT20G catalog data.
It is possible that this source is variable.

For J040446-294011, the flux densities from the VLASS and AT20G catalogs are well approximated by a log-parabolic function in the range of 3--20 GHz and a log-linear function \footnote {The log-linear function or linear function on a logarithmic scale is $log (S_{\nu}) = a log (\nu) +c.$} with a spectral index $\alpha$ = --0.49, which indicates the presence of extended non-thermal radio emission from parts of the radio source.
However, if we assume that the flux density "bounce" at 3 GHz is due to the variability of the source, then the spectral index $\alpha_{below}$ changes from 1.519 to 0.677.

The examples given show that we cannot always confidently approximate the spectrum of the source. In this regard, in table~\ref{tab:Sp.45+} for a number of sources two variants of spectrum parameters are given, or only flux density values from the AT20G catalog are used when constructing the spectrum. Comments are given in column (12) of table~\ref{tab:Sp.45+}.

Analyzing the parameters of the spectra of the radio sources in the sample, we can state that all of them, with the exception of six sources, satisfy the HFP criterion -- their peak frequencies are $\nu_{obs}$ > 5 GHz. Of these six sources, four have $\nu_{obs}\simeq$5\,GHz, and only two sources, J002442-420203 and J155941-244240, do not satisfy the $\nu_{int}$>5\,GHz condition in the source rest frame.

The figure~\ref{fig1} shows histograms of the distributions of the spectral indices $\alpha_{below}$ and $\alpha_{above}$, as well as FWHM, $S_{peak}$, $\nu_{obs}$, and $\nu_{int}$. The values of the means and medians for the spectral parameters of the radio sources are given in the table~\ref{tab:Sp1}. When constructing the histograms and estimating the values of the means and medians, planetary nebulae, HII regions, stars, and young stellar objects were not taken into account.

Note that 67\% of our sample are radio sources for which there are no data on flux densities in the low-frequency region of the spectrum, namely at the frequencies of the GLEAM and TGSS catalogs (group 1). The average values of the spectral indices $\alpha_{below}$ of the sources in this group are 30\% greater than those of the sources for which there are data in the GLEAM and TGSS catalogs (group 2). The number of sources in this group with a spectral index $\alpha_{below}$ > +1 is an order of magnitude greater than the number of sources in the second group with the same $\alpha_{below}$ values.

\begin{table}
\caption{
Average and median values of the parameters of the spectra of radio sources of the formed sample.
}
\begin{tabular}{|l|l|l|}
\hline\hline
Parameter	     &~~~Mean & Median \\
\hline
~$\alpha_{below}$ &~~0.85$\pm$0.31 &~~~0.76    \\
~$\alpha_{above}$ &~-0.49$\pm$0.19 &~~-0.48    \\
~FWHM             &~~1.40$\pm$0.11 &~~~1.40    \\
~$\nu_{obs}$ (GHz) &~~17.6$\pm$14.3 &~~~12.9    \\
~$\nu_{int}$ (GHz) &~~34.8$\pm$30.7 &~~~26.1    \\
~$S_{peak}$ (Jy)   &~~0.21$\pm$0.30 &~~~0.12    \\
\hline\hline
\end{tabular}
\label{tab:Sp1}
\end{table}
The average values of $S_{peak}$, $\nu_{obs}$, and $\nu_{int}$ for the first group of sources are 24\%, 37\%, and 76\% lower than the average values of the same parameters for the second group of sources, respectively.
Both groups contain objects with purely inverted spectra. Such radio sources most likely have spectral maxima at frequencies above 20 GHz.
The number of sources with such spectra is about two dozen in each group.

The average values of the spectral peak flux densities for the second group of sources are higher than for the first group -- 0.30$\pm$0.35 Jy and 0.17$\pm$0.26 Jy, respectively. Some of them have flux densities greater than 1 Jy.

\section{GALAXIES AND QUASARS IN THE SAMPLE}
In high spatial resolution images, HFP quasars have complex asymmetric or core-jet morphologies. In most cases, the core-jet morphology at parsec scales indicates that the source is a blazar. GPS galaxies usually have a two- or three-component structure and are classified as compact symmetric objects (CSO).
In some cases, a weak flat component is detected in the spectrum, which is interpreted as emission from the core. 

Note that in \cite{2008A&A...482..483T}, where clustering algorithms were used to separate the GPS features, no clusters were found in the case of the morphological classification, indicating the existence of distinct subpopulations in addition to the expected separation of galaxies and quasars. This was expected if one follows the unified AGN model \citep{1989ApJ...336..606B, 1995PASP..107..803U}, where one of the observed differences between radio galaxies and radio quasars is the angle at which the jet and active nucleus are located to the observer's line of sight.

In this section, we analyze the parameters of the continuum spectra of galaxies and quasars in our sample and compare their properties.

The main criterion for forming our sample of sources from the AT20G catalog is the restriction on the spectral index: $\alpha_{below}\ge$+0.5. In the sample formed in this way, the number of quasars is almost 2.5 times greater than the number of galaxies. Moreover, of the total number of sources in the sample (255), the proportion of quasars is 71\%, and galaxies 28\%. These values are close to the ratio of quasars and galaxies -- 77\% and 23\%, respectively, in the sample that was formed based on the restriction on the peak frequency $\nu_{obs}$>5\,GHz in \cite{2009AN....330..180H}.

Note that our sample included 66 blazars and BLLac objects. Some of them are not always reliably identified in the literature; in Table~\ref{tab:Sp1} they are designated as "Bz?".

This ratio between the number of galaxies and quasars in the GPS/HFP candidate samples is consistent with the trend noted in \cite{1990A&A...231..333F, 1998PASP..110..493O, 2005A&A...432...31T, 1998A&AS..131..303S, 2003PASA...20..118S} towards a decreasing number of galaxies relative to the number of quasars with increasing peak frequencies. At the same time, the proportion of blazars in the sample will also increase with increasing peak frequency. Young objects become increasingly rare as their lifetimes decrease with increasing peak frequency.

\begin{figure}
\centerline{
\vbox{
\includegraphics[angle=0,width=0.33\textwidth,clip]{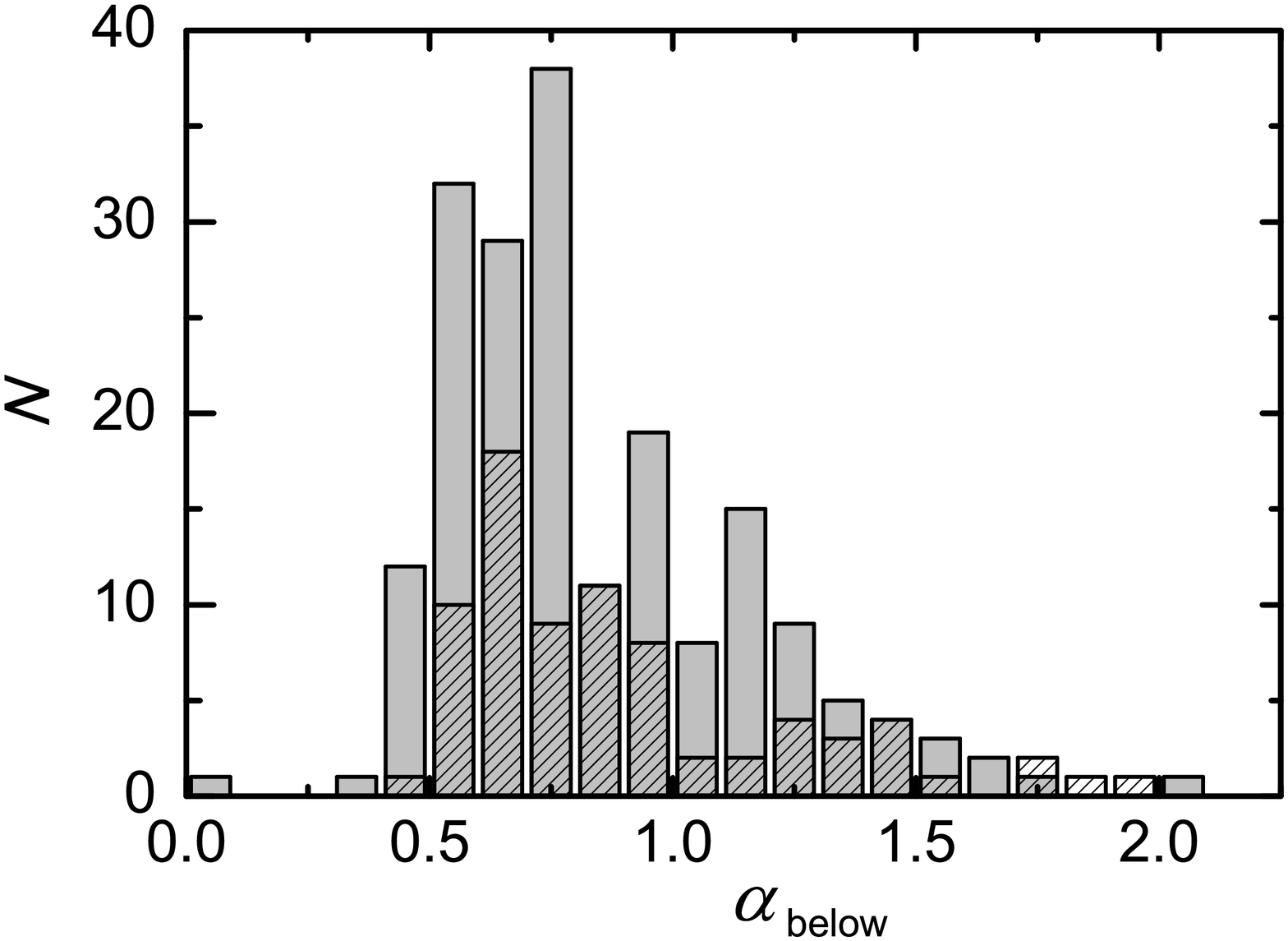}
\includegraphics[angle=0,width=0.33\textwidth,clip]{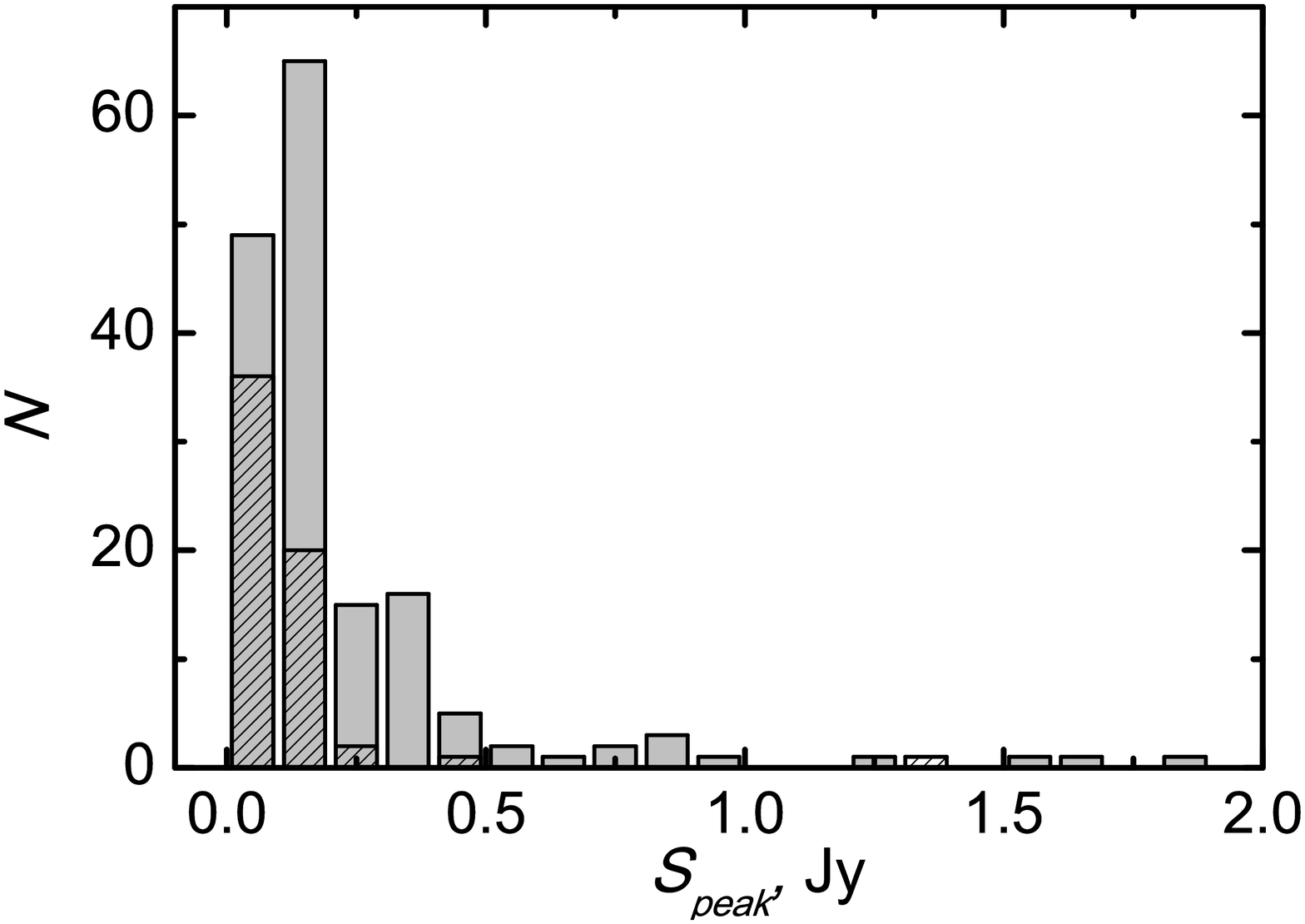}
\includegraphics[angle=0,width=0.33\textwidth,clip]{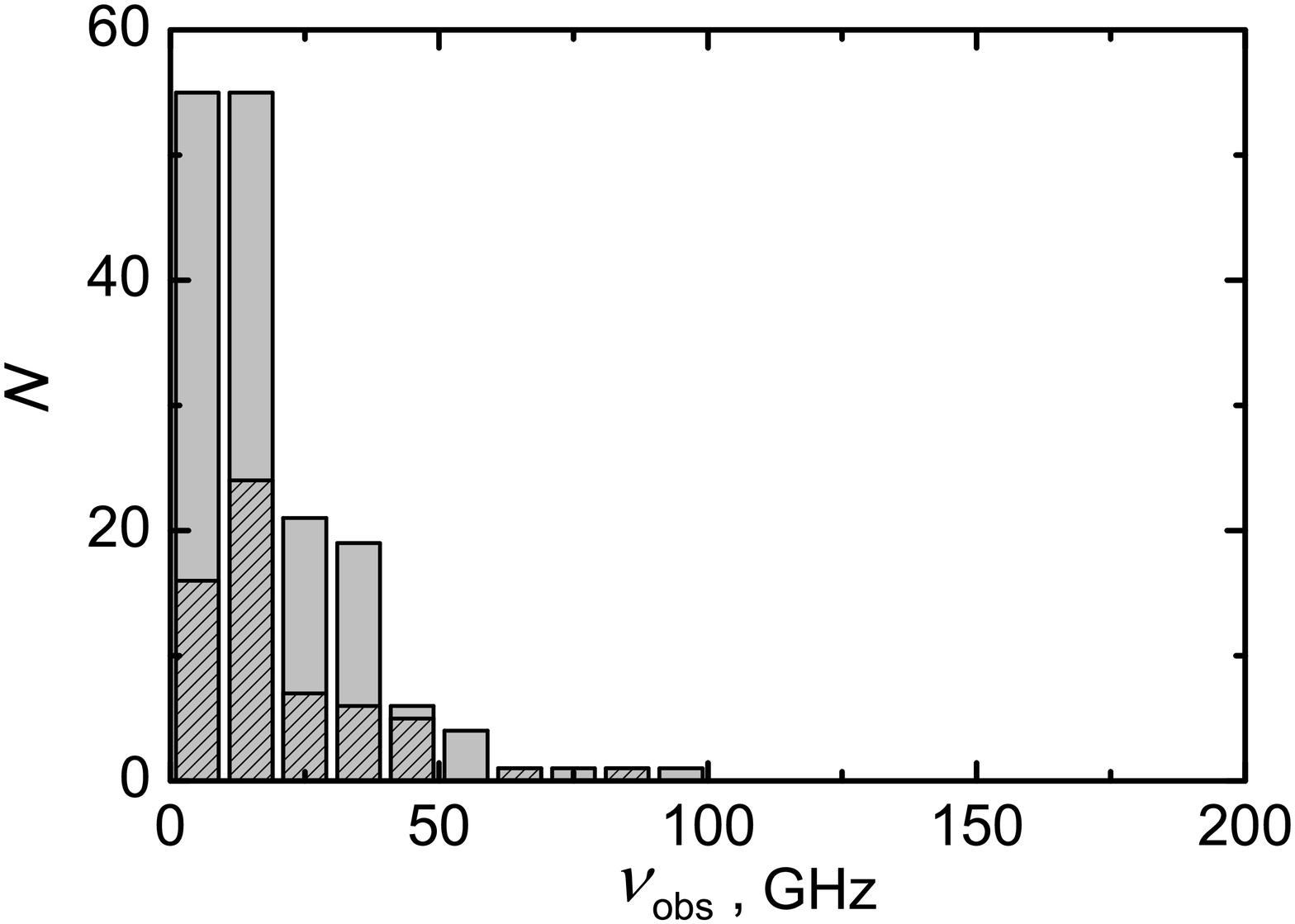}
\includegraphics[angle=0,width=0.33\textwidth,clip]{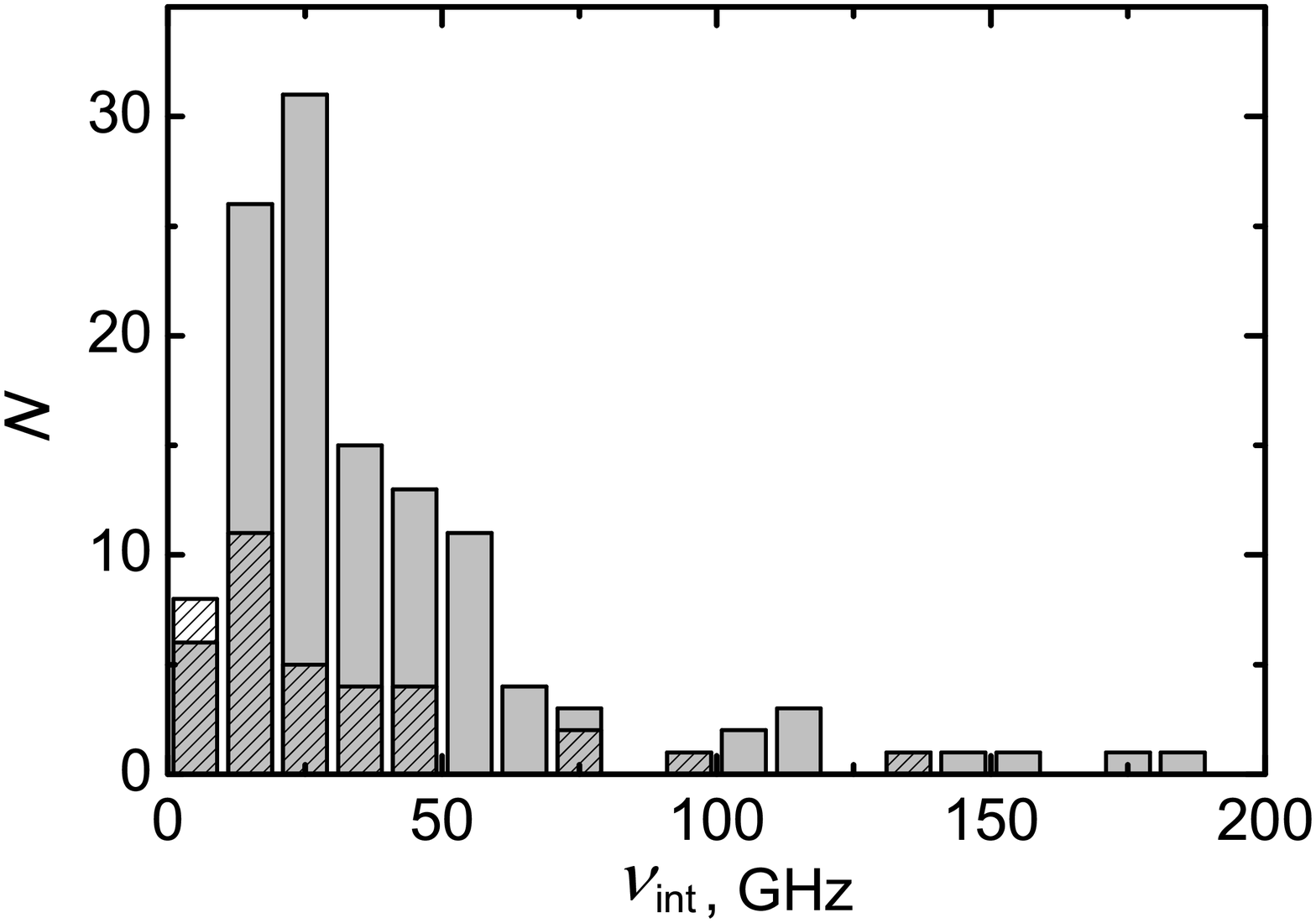}
}
}
\caption{
Histograms of the distributions of spectral indices $\alpha_{below}$, peak flux densities $S_{peak}$, peak frequencies in the observer coordinate system $\nu_{obs}$ and peak frequencies in the source coordinate system $\nu_{int}$ for quasars (gray rectangles) and galaxies (shaded rectangles) of the source sample.
}
\label{fig4}
\end{figure}

Figure~\ref{fig4} shows histograms of the distributions of the following parameters $\alpha_{below}$, $S_{peak}$, $\nu_{obs}$ and $\nu_{int}$ for the galaxies and quasars in our sample. Gray rectangles correspond to the spectral parameters of quasars, and shaded rectangles correspond to galaxies. The mean and median values of the parameters are given in Table~\ref{tab:GQ1}.

The $\alpha_{below}$ distributions for galaxies and quasars turned out to be similar.
The maximum of the Gaussian fitted into the $\alpha_{below}$ distribution for galaxies was 0.73$\pm$0.02, half-width 0.35, for quasars these values were 0.70$\pm$0.03 and 0.41, respectively.

The average $\overline{\alpha_{below}}$ values for galaxies are approximately 8\% higher than for quasars. In \cite{2013AstBu..68..262M} this difference was $\sim $ 10\%.

\begin{figure}
\centerline{
\includegraphics[angle=0,width=0.35\textwidth,clip]{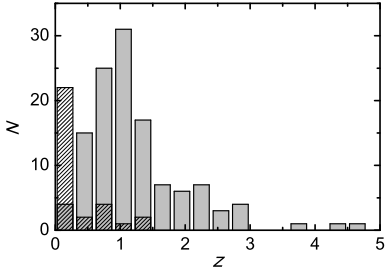}
}
\caption{
Histogram of the redshift distributions of quasars (gray rectangles) and galaxies (shaded rectangles) from our sample.
}
\label{fig5}
\end{figure}

\begin{figure}
\centerline{
\includegraphics[angle=0,width=0.47\textwidth,clip]{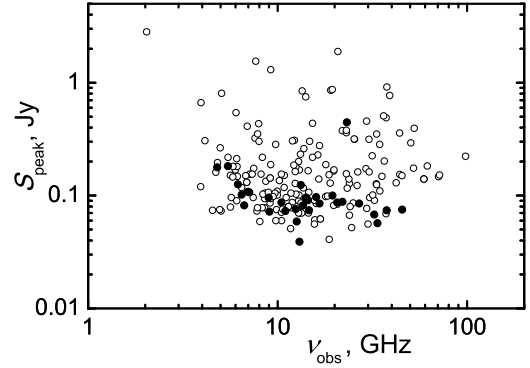}
}
\caption{
Dependence of the peak flux density $S_{peak}$ on the frequency $\nu_{obs}$. Filled black circles correspond to galaxies, unfilled ones to quasars.
}
\label{fig6}
\end{figure}

The average and median values of the $\nu_{obs}$ frequency distributions for peaks in the spectra of galaxies and quasars practically coincided. The average peak frequencies of $\nu_{int}$ quasars are significantly higher than those of galaxies -- 40.4$\pm$34.0 GHz and 28.9$\pm$27.9 GHz, respectively, which is explained by the large redshifts of quasars compared to galaxies. Histograms of the redshift distributions of quasars and galaxies are shown in the figure~\ref{fig5} (gray rectangles refer to quasars, shaded ones to galaxies).
The average $z$ values of quasars were 1.24, galaxies — 0.31, the medians — 1.00 and 0.15, respectively.
\begin{table}
\caption{
Average and median values of spectral parameters of galaxies and quasars.
}
\begin{tabular}{|l|l|l|l|l|}
\hline\hline	    
~Param. &~Mean Q & Mean G & Med. Q & Med. G  \\
\hline
~$\alpha_{below}$ & 0.83$\pm$0.31 &~0.90$\pm$0.36 &~0.76  &~0.80    \\
~$\alpha_{above}$ & -0.53$\pm$0.20 &~-0.43$\pm$0.15 &~-0.52 &~-0.49 \\
~FWHM             & 1.42$\pm$0.11 &~1.36$\pm$0.09 &~1.40 &~1.20    \\
~$\nu_{obs}$      & 19.1$\pm$15.5 &~20.0$\pm$15.7 &~13.2 &~14.1    \\
~$\nu_{int}$      &~40.4$\pm$34.0 &~28.9$\pm$27.9 &~28.8 &~18.1    \\
~$S_{peak}$       &~0.24$\pm$0.34 &~0.13$\pm$0.17 &~0.14 &~0.09    \\
\hline\hline
\end{tabular}
\label{tab:GQ1}
\end{table}
As for the flux densities of the sources, the vast majority of galaxies have flux densities <0.2\,Jy, quasars have flux densities <0.4\,Jy, and the average values of $\overline{S_{peak}}$ are 0.13\,Jy and 0.24,Jy, respectively (table~\ref{tab:GQ1}). The fact that the flux densities $S_{peak}$ of galaxies and quasars differ significantly is clearly demonstrated by Fig.~\ref{fig6}, which shows the dependences of the peak flux density $S_{peak}$ on the peak frequency $\nu_{obs}$. Filled black circles denote galaxies, unfilled ones denote quasars.
In \cite{2010MNRAS.408.1075O} it was suggested that young sources would be more common among "faint" galaxies with flux densities < 0.3 Jy. Thus, it can be assumed that young HFP radio sources may be present among the galaxies in our sample.

\begin{figure}
\centerline{
\includegraphics[angle=0,width=0.47\textwidth,clip]{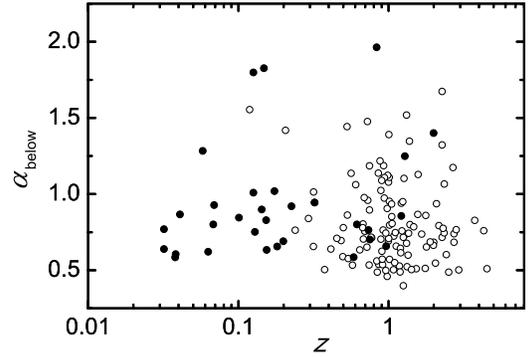}
}
\caption{
Dependence of $\alpha_{below}$ on the redshift $z$. Black filled circles indicate galaxies, unfilled circles indicate quasars.
}
\label{fig7}
\end{figure}

The figure~\ref{fig7} shows the dependence of $\alpha_{below}$ on $z$. The open circles indicate the $\alpha_{below}$ values of sources identified with quasars, and the filled circles indicate galaxies.
The graph presented in Fig.~\ref{fig7} is more informative than the histograms of the $\alpha_{below}$ and $z$ distributions and complements them.
Thus, most galaxies have a spectral index $\alpha_{below}$<+1.1 and a redshift $z$<0.25.
A clear boundary is traced at z$\sim$0.4, to the left of which are located most galaxies, to the right -- most quasars. The redshifts of quasars lie in the range 0.4<$z$<5.

The view of the graph in Fig.~\ref{fig7} is most likely associated with observational selection. In optics, galaxies are weaker than quasars, and for this reason, the literature currently mainly presents galaxies with a redshift z<1.

It can be noted that the vast majority of galaxies in our sample belong to the first group of sources, for which there are no data at frequencies below 0.8 GHz.
And quasars mainly belong to the second group of radio sources, for which there are data on flux densities at low frequencies.
This group includes only 10 galaxies, which is an order of magnitude less than quasars. Most likely, this is due to the fact that galaxies have lower flux densities then quasars and are not detected at low frequencies.

In conclusion, let us compare some characteristics of our sample with similar characteristics of the sample from \cite{2009AN....330..180H}. As already noted, in this paper the sample of 656 GPS sources from the AT20G catalog was formed based on the frequency constraint on the peak in the spectrum $\nu_{obs}$ > 5 GHz. Our sample, including 255 sources, was also formed based on the AT20G catalog, but according to a different selection criterion, namely $\alpha_{below}\ge$+0.5.

The ratio of quasars and galaxies in our sample (71\% and 28\%) practically coincided with the ratio of quasars and galaxies (77\% and 23\%) in the sample from \cite{2009AN....330..180H}. The mean redshift z values of quasars are 1.24 and 0.31 for galaxies, while in \cite{2009AN....330..180H} these values are 1.2 and 0.2, respectively. As noted in \cite{2009AN....330..180H}, "galaxies have a sharper peak around 0.1--0.35 Jy of the inferred peak flux, while quasars have a broader peak around 0.35--1 Jy". The flux densities $S_{peak}$ of galaxies and quasars in our sample are 0.13$\pm0.17$ Jy and 0.24$\pm0.34$ Jy, respectively. The peak frequencies of the $\nu_{obs}$ sources in \cite{2009AN....330..180H} lie in the range of 8--15 GHz, while most of the sources in our sample have $\nu_{obs}$ in the range of 5--50 GHz.

\section{Sources with ultra-inverted spectra}

\citep{2010MNRAS.405.1560M} contains a list of radio sources that belong to a small but important group of non-thermal sources with spectral indices $\alpha^{20}_{5}$ > +0.7.
By analogy with radio sources with ultra-steep spectra $\alpha$<--1.3 \citep{1979A&A....80...13B,1979A&AS...35..153T} the term "ultra-inverted spectrum" (UIS) was introduced for these sources. It was suggested that the steep spectrum of these sources, caused by self-absorption, indicates their youth or, perhaps, that they are embedded in a very dense medium \citep{2010MNRAS.405.1560M}. If the peak in the source spectrum is at frequencies $\geq$ 20 GHz, then the ratio between the spectral peak and the source size yields a size of less than a few tens of parsecs. It is estimated that 1.2\% of AT20G sources will belong to this class.

Among the radio sources in our sample, 187 sources with $\alpha_{below}\geq$+0.7 were found. This is 3.2\% of the sources in the AT20G catalog and 70\% of the sources in our sample.
Of these, 42 sources have data at the frequencies of the TGSS and GLEAM catalogs. If there are young sources among them, then they most likely belong to the restarted ones. To speak about this more definitely, additional studies are needed.

Some of the radio sources with ultra-inverted spectra from Tables 8 and 9 in \cite{2010MNRAS.405.1560M} were also included in our sample when selecting sources with $\alpha_{below}\geq$+0.5. In Table~\ref{tab:Sp.45+} (column 12) these sources are designated by the symbols t8 (from Table 8), t9 (from Table 9), t89 (from Tables 8 and 9).

Table 8 \citep{2010MNRAS.405.1560M} lists AT20G sources that have no analogues in the SUMSS or NVSS catalogs.
24 sources listed in this table have spectral index $\alpha^{1}_{20}$>+0.7. We constructed their spectra using CATS data. It turned out that 16 sources have $\alpha_{below}$>+0.7, and 8 sources have $\alpha_{below}$<+0.7.

Table 9 \citep{2010MNRAS.405.1560M} lists AT20G sources with ultra-inverted spectra with $\alpha^{5}_{20}\ge$+0.7. Of these 45 sources, 24 have $\alpha_{below}\ge$+0.7. Eight of these are already listed in Table 8.

Thus, of the AT20G sources listed in Tables 8 and 9 \citep{2010MNRAS.405.1560M}, 69 sources have spectral indices $\alpha^{1}_{20}$>+0.7 and $\alpha^{5}_{20}\ge$+0.7.

Note that after constructing the spectra of these sources using CATS data and calculating the spectral indices, only 32 sources were confirmed to have a spectral index that satisfies the condition $\alpha_{below}\ge$+0.7. We believe it is important to use the entire available set of flux densities at different frequencies when constructing the radio spectrum, since this allows us to more reliably determine the spectral index.

In \cite{2011MNRAS.412..318M} four sources with spectral indices $\alpha^{5}_{8}$>+2.5 were detected. As noted in this paper, they correspond to the highest possible synchrotron spectral self-absorption indices of +2.5 at low frequencies and are inconsistent with the maximum free-free absorption spectrum of +2.0. Of these four sources, three were included in our sample: J070949-381152, J111246-203932, and J143608-153609 (see Table~\ref{tab:Sp.45+}). The spectral indices $\alpha_{below}$ of these sources were less than +2.0 and were 1.80, 1.79, and 1.0, respectively.

The most probable HFP candidates are considered to be radio sources with ultra-inverted spectra with a narrow peak in the spectrum, which was noted in early HFP studies \citep{1991ApJ...380...66O, 1997A&A...321..105D, 2004A&A...424...91E}, and for such sources FWHM = 1.2 frequency decades was adopted. 

According to the results of later studies \citep{2013AstBu..68..262M, 2012A&A...544A..25M} the FWHM value turned out to be larger: from 1.4 to 1.5. As a boundary for selecting the most probable HFPs, we adopted the FWHM value = 1.35, close to the median value of our sample. In the table~\ref{tab:Sp.45+} the names of sources that satisfy the conditions $\alpha_{below} \geq +0.7$ and FWHM $ \leq$ 1.35 are highlighted in bold.
Among the sources in our sample with measured spectral half-widths, 45 were found that satisfy these conditions.
Of these, 30 were identified with quasars, 11 with galaxies, 3 with planetary nebulae, and one was not identified.
The values of the mean and median spectral parameters for these sources are given in Table~\ref{tab:Sp2}. Planetary nebulae were not used in the estimates of the mean and median spectral parameters.
\begin{table}
\caption{
Average and median values of spectral parameters of sources with $\alpha_{below} \geq +0.7$ and FWHM $ \leq$ 1.35.
}
\begin{tabular}{|l|l|l|}
\hline\hline
Parametr &~Mean & Median        \\
\hline\hline
~$\alpha_{below}$ &~1.10$\pm$0.26 ~&~~~1.10   \\
~$\alpha_{above}$ &-0.56$\pm$0.17 ~&~~-0.58    \\
~FWHM             &~1.30$\pm$0.04 ~&~~~1.30   \\
~$\nu_{obs}$ &~ 9.9$\pm$7.9  ~&~~~ 7.9   \\
~$\nu_{int}$ &~23.1$\pm$20.7 ~&~~~15.9   \\
~$S_{peak}$  &~0.20$\pm$0.22 ~&~~~0.15   \\
\hline\hline
\end{tabular}
\label{tab:Sp2}
\end{table}

The mean and median values of the spectral indices of the $\alpha_{below}$ sources in this subsample, as expected, were higher than those of the full sample of sources, and the FWHM values were lower. The mean values of the peak flux densities were almost identical: 0.21$\pm$0.30 Jy and 0.20$\pm$0.22 Jy, respectively.

The mean and median values of the peak frequencies $\nu_{int}$ and $\nu_{obs}$ of the subsample of sources with ultra-inverted spectra were 50\% and 70\% lower than those of the full sample. The subsample did not include radio sources with a peak at frequencies $\nu_{obs}$ > 50 GHz and $\nu_{int}$ > 100 GHz.
However, if we remove the restriction on the half-width of the spectrum $FWHM\leq$1.35, then the sample will include several sources with $\nu_{obs}$>50 GHz, as well as sources with inverted spectra without a peak, for which the peak frequencies $\nu_{obs}$ can be higher than 50 GHz.

\section{Radio luminosity of galaxies and quasars}

For radio sources in the sample with known redshifts, the radio luminosity $L_{20}$ at 20 GHz was calculated using the formula from \cite{2008A&ARv..15...67M}: \\
$L_{\nu}=4\pi D_{L}^{2}S_{\nu}(1+z)^{-(\alpha+1)}$, (1) \\
where $S_{\nu}$ is the source flux density at $\nu$, $D_{L}$ is the luminosity distance.
We used $\Lambda$CDM cosmology with $H_{0}$=67.4 km/sec/Mpc, $\Omega_{m}$ = 0.315, and $\Omega_{\Lambda}$ = 0.685 \citep{2020A&A...641A...6P}.
The $D_{L}$ values were calculated using the cosmology module of the astropy package (https://www.astropy.org/).

The radio luminosity was calculated only for sources whose flux densities at 20 GHz were in the spectral region characterizing the optically thin region of emission, and for which the spectral indices $\alpha_{above}$ were determined.
The radio luminosity values $L_{20}$ are given in Table~\ref{tab:Sp.45+} (column 9).

\begin{figure}
\centerline{
\includegraphics[angle=0,width=0.47\textwidth,clip]{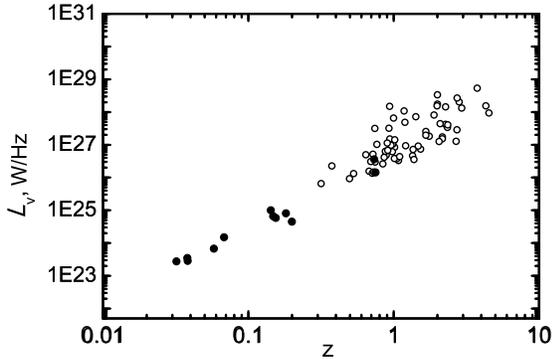}
}
\caption{
Dependence of radio luminosity $L_{\nu}$ of sources at 20 GHz on redshift $z$. Unfilled circles correspond to quasars, filled circles to galaxies.
}
\label{fig9}
\end{figure}

Fig.~\ref{fig9} shows the dependence of the radio luminosity $L_{\nu}$ on the redshift $z$ for our sample. Filled black circles denote galaxies, unfilled ones denote quasars.
The radio luminosities of galaxies lie in the range $10^{23}-10^{26}$ W/Hz, and those of quasars — $10^{26}-10^{30}$ W/Hz.
The ranges of radio luminosities of galaxies and quasars coincided with the range of radio luminosities of sources from the AT20G catalog at a frequency of 20 GHz (see \cite{2011MNRAS.417.2651M}), and the nature of the dependence of $L_{20}$ on $z$ also coincided.

\section{Estimates of angular sizes of emitting regions of the sample galaxies}

Estimates of the upper limits of the angular sizes of the emitting regions of the galaxies from our sample were made. Estimates are made under the assumption of a radio source uniform in structure and magnetic field with a power-law distribution of emitting particles and self-absorption at frequencies below the peak frequency. Since the condition of equidistribution of energy between the magnetic field and relativistic particles in the jet of quasars/blazars is not satisfied, we limited ourselves to estimates of the angular sizes of the emitting regions of galaxies.
For them, the frequency of the spectrum maximum is related to $\theta$ by the following relation \citep{2000MNRAS.319..445S, 1981ARA&A..19..373K, 2013AstBu..68..262M, 2019AstBu..74..348S}: \\
$\nu_{max}=8B^{1/5}S_{max}^{2/5}\theta^{-4/5} (1+z)^{1/5}$, (2) \\
where B is the magnetic field in gauss, $S_{max}$ is the flux density at the maximum of the radio spectrum in Jy, $\nu_{max}$ is the frequency of the maximum in the spectrum in GHz. In the previously adopted notations, $S_{max}$ corresponds to $S_{peak}$, $\nu_{max}$ — to the frequency of the peak $\nu_{obs}$. For compact extragalactic objects with a uniform distribution of the magnetic field and relativistic particles, the magnetic field B is taken to be 100 $\mu$G \citep{1985ApJ...290...86M}\footnote{Note that $\theta$ weakly depends on B.}.
Then $\theta$ can be calculated by the formula: \\
$\theta\approx1.345\sqrt{S_{max}}(1+z)^{1/4}\nu_{max}^{-5/4}$. (3) \\
The angular sizes $\theta$ of the galaxies in the sample are presented in the table~\ref{tab:Sp.45+} (column 8).

\begin{figure}
\centerline{
\includegraphics[angle=0,width=0.4\textwidth,clip]{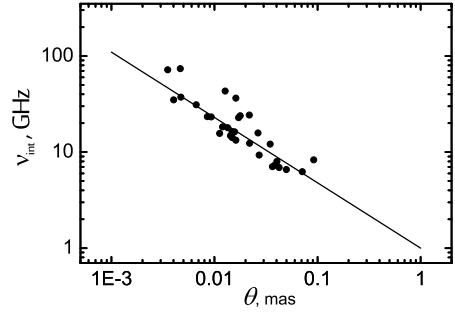}
}
\caption{
Dependence of the peak frequency $\nu_{int}$ on the angular sizes of the galaxies $\theta$ in the sample. The black line shows the dependence $\nu_{int}=\theta^{-0.68}$ on a logarithmic scale.
}
\label{fig11}
\end{figure}

The dependence of $\theta$ on $\nu_{int}$ shown in Fig.~\ref{fig11} demonstrates that there is an anticorrelation between the angular sizes of galaxies $\theta$ and their own peak frequencies $\nu_{int}$, which was previously discovered in CSS and GPS radio source observations
in \cite{1990A&A...231..333F,1990cssg.conf...55S}. The values of $\theta$ and $\nu_{int}$ are presented in logarithmic scale.
The black line in Fig.~\ref{fig11} shows the dependence $\nu_{int} = \theta^{-0.68}\,$ in logarithmic scale.
The behavior of the frequencies of the peaks in the radio spectrum for galaxies in the source reference frame $\nu_{int}$ on their angular sizes $\theta$ is in good agreement with the regression $\nu_{int}\simeq\theta^{-0.68}$ obtained from the relation:\\
$log(\nu_{int})=(-0.68\pm0.01)\,log(\theta)$.

The dependence of $\theta$ on $\nu_{int}$ in Fig.~\ref{fig11} is close to that presented in \cite{2019AstBu..74..348S}.
The difference is that the sources in our sample with known redshifts, with the exception of one, have peak frequencies $\nu_{int}$ above 5 GHz and, accordingly, smaller angular sizes, down to 0.003 mas.

A similar anticorrelation was found in \cite{1990A&A...231..333F, 1998PASP..110..493O, 2014MNRAS.438..463O} between the linear sizes of LS sources and their $\nu_{int}$.

In \cite{2014MNRAS.438..463O}, an analytical expression for the relationship between the peak frequency $\nu_{int}$ and the largest linear size of sources is given:\\
$log(\nu_{int})=-0.21\pm0.04-(0.59\pm 0.05)\,log(LS)$.\\
It was obtained from an approximation of the experimental dependence of LS on $\nu_{int}$, constructed on the basis of observational data.
A simplified formula is often used for estimates: $\nu_{int} \simeq LS^{-0.59}$.
In \cite{1998PASP..110..493O} the relationship between LS and $\nu_{int}$ is expressed by a similar formula:\\
$log(\nu_{int})=-0.21\pm0.05-(0.65\pm0.05)\,log(LS)$ or $\nu_{int}\simeq LS^{-0.65}$.\\
Using the relationship between the peak frequency and the linear size (LS) of the source, we estimated the linear sizes of the sources in our sample.
The linear sizes of the sources are small and range from 0.2 pc to 30 pc for galaxies, less than 30 pc, and for most quasars, less than 10 pc. The fact that high-frequency GPS sources are very compact ($\sim10\div 50$ pc) has already been noted earlier, for example, in \cite{2010MNRAS.408.1187H, 1998PASP..110..493O, 1997AJ....113..148O}. 

Thus, our sample consists of compact radio sources with high peak frequencies (>5 GHz), small angular and linear sizes, among which quasars predominate.
Since, according to the "youth" scenario, there is an anticorrelation between the source size and its age \cite{2000MNRAS.319..445S, 1995A&A...302..317F}, caused by synchrotron self-absorption, our sample may contain radio sources with an age of $10^{2}-10^{3}$ years \citep{2003astro.ph..9354T}.

However, the search for such sources is complicated by the fact that their lifetime is short. They either decay before reaching large sizes, or their peak frequencies shift to lower frequencies as they age. Another problem is contamination of such samples by blazars, which can be very compact due to the orientation of their jets toward the observer, and also exhibit inverted or peaked spectra during outbursts. Young radio sources differ from blazars in that they are considered the least variable class of extragalactic objects and are practically unpolarized, while blazars exhibit significant spectral variability and their emission is polarized \cite{2006A&A...450..959O}.

\section{Magnitude-Redshift Relationship}

As shown by numerous studies, a significant proportion of quasars with a high-frequency peak in the spectrum are actually flaring sources with a flat spectrum or blazars. In our sample, blazars accounted for 24.5
Almost the same percentage (25

It is known that true HFP quasars are quite rare, and their detection requires long-term monitoring of their spectrum. Among galaxies, true HFPs are much more common \cite{2007A&A...469..451T}.

We define true HFP sources as those which, in accordance with the characteristics given in \cite{2010MNRAS.408.1187H}, in addition to a convex radio spectrum with a peak above 5 GHz, weak flux density variability and small variability of the spectrum shape over time, have a compact radio morphology and an optical spectrum with broad or narrow emission or absorption lines, rather than a spectrum without continuum features, as is observed in blazars. Such characteristics should be met, in particular, by young radio galaxies with high-frequency peaks in the radio spectrum \cite{2010MNRAS.408.1187H}.

\cite{1998PASP..110..493O, 1996MNRAS.279.1294S, 2002MNRAS.337..981S} showed that for GPS galaxies the "$R$ band  magnitude -- redshift"\ dependence agrees well with the Hubble relation.
Later, M.~Orienti et al. \cite{2010MNRAS.408.1075O} showed that for HFP galaxies the "$R$ -- z"\ dependence also agrees well with the Hubble relation using a sample of HFP galaxies from \cite{2009AN....330..223S}.

\begin{figure}
\centerline{
\vbox{
\includegraphics[angle=0,width=0.45\textwidth,clip]{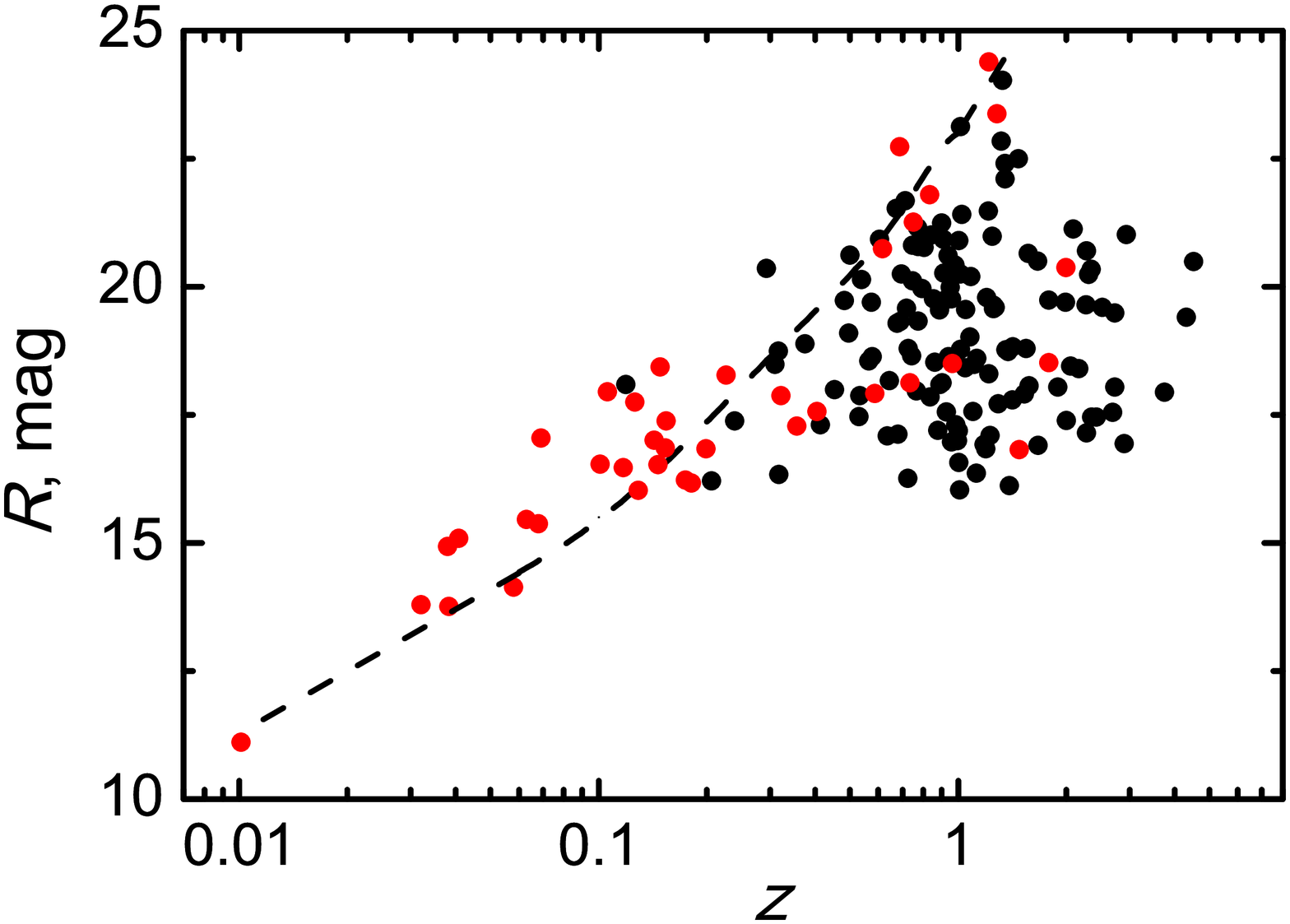}
\includegraphics[angle=0,width=0.45\textwidth,clip]{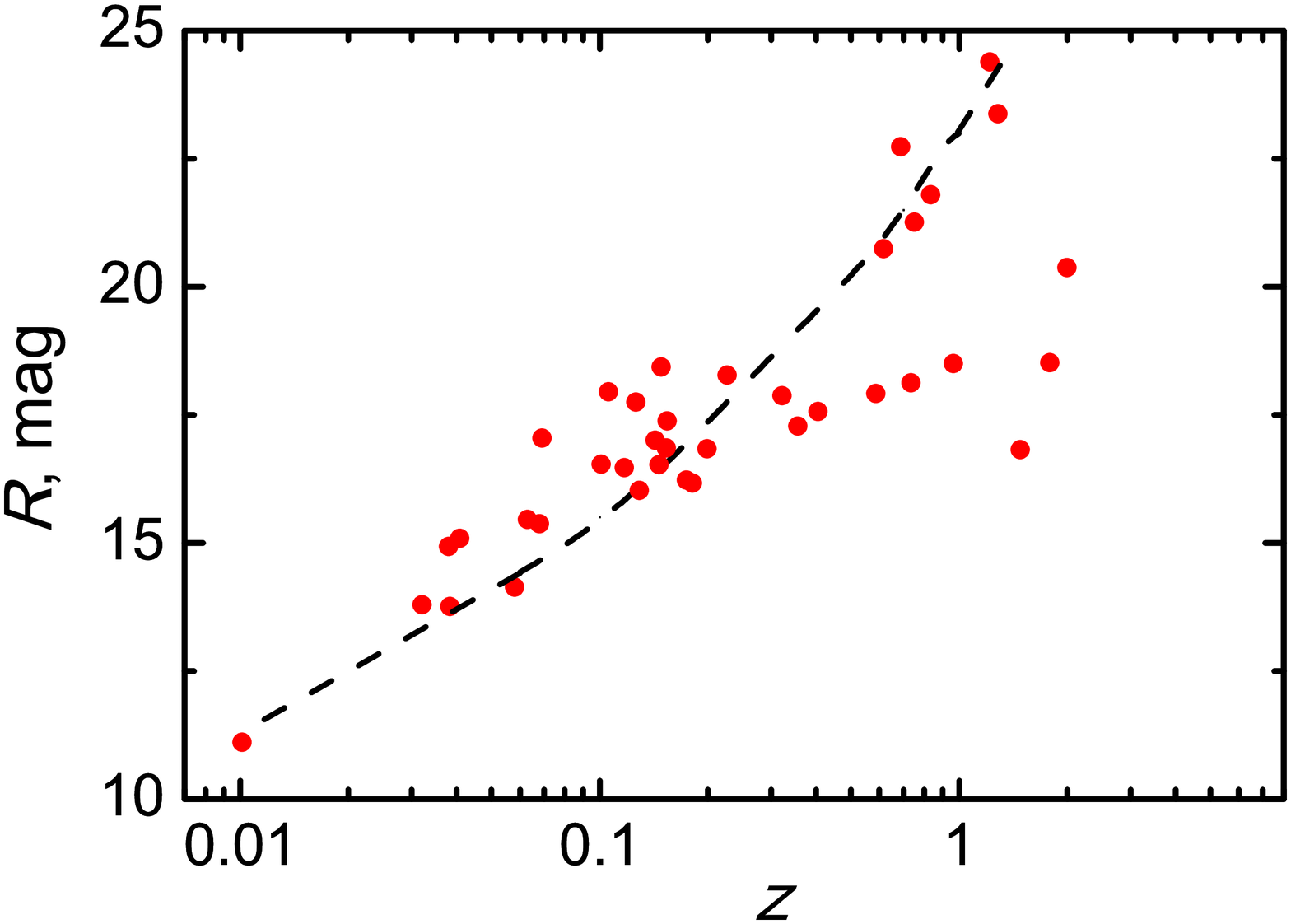}
}
}
\caption{
The left panel shows the dependence of the stellar magnitude in the $R$ band on the redshift $z$ for the galaxies and quasars of our sample, the right panel shows only the galaxies.
Galaxies are shown as red filled circles, quasars as black ones. The dotted line shows the Hubble relation \citep{1996MNRAS.279.1294S}.
}
\label{fig13}
\end{figure}

To check whether the galaxies in our sample obey the Hubble relation \cite{1996MNRAS.279.1294S}, "$R$ -- $z$" dependencies were constructed. The dependencies were constructed both for all radio sources in the sample (Fig.~\ref{fig13}, left panel) and only for radio galaxies (Fig.~\ref{fig13}, right panel).
Red filled circles denote galaxies, black ones -- quasars. The dotted line shows the Hubble relation.
It is evident from the presented dependences that for some galaxies their stellar magnitudes deviate quite strongly from the Hubble dependence.

\begin{figure}
\centerline{
\includegraphics[angle=0,width=0.45\textwidth,clip]{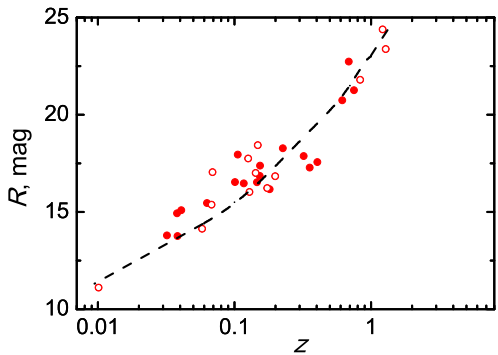}
}
\caption{
Dependence of stellar magnitude $R$ on redshift $z$ for galaxies with the smallest deviations of $R$ from the Hubble relation.
The empty red circles indicate $R$ of galaxies for which $\alpha_{below} \geq 0.7$ and FWHM $\le$ 1.35. The dotted line shows the Hubble relation \citep{1996MNRAS.279.1294S}.
}
\label{fig15}
\end{figure}

The table~\ref{tab:noHabbl} lists radio sources identified with galaxies with the largest deviations of $R$ values from the Hubble dependence.
Galaxies for which these deviations are small (< $\pm$ 2.5 mag.) are listed in the table~\ref{tab:yesHabbl}, and the $R$ -- $z$ dependences for them are shown in Fig.~\ref{fig15}. The unfilled red circles in the figure indicate galaxies for which $\alpha_{below} \geq 0.7$ and FWHM $\le$ 1.35.
Information about sources in Table~\ref{tab:noHabbl} and \ref{tab:yesHabbl}
partially coincides with that given in Table~\ref{tab:Sp.45+}. Additionally, in
columns (8) the stellar magnitudes $R$ of sources in the R band from the optical catalog AT20G are given. The catalog is available in the electronic
version of the article \cite{2011MNRAS.417.2651M}, http://ssa.roe.ac.uk//.
Columns 12 present information about the optical identification of sources from the optical catalog AT20G \cite{2011MNRAS.417.2651M}.
\begin{table*}
\caption{
Parameters galaxies with the largest deviations of magnitudes $R$ from the Hubble relation: $\alpha_{below}$, FWHM, $\nu_{obs}$, $S_{peak}$, angular sizes $\theta$, radio luminosities $L_{\nu}$ at 20 GHz, redshifts $z$, magnitudes in the $R$ band, the type of radio source TypeR and the type of parent object Type, optical classification of objects from \cite{2011MNRAS.417.2651M}.
}
\begin{tabular}{|c|l|l|l|l|l|l|l|l|l|l|l|}
\hline\hline
 ~~NVSS      ~            &$\alpha_{below}$ &~FWHM         &~~$\nu_{obs}$  &~$S_{peak}$ &~~~~~$\theta$  &~$L_{\nu}$   &~~~$R$     &~~$z$          &~Type  &~~TypeR   &~~opt.    \\
 ~~name      ~            &                 &~~            &~GHz &~~Jy &~mas &~W/Hz &~mag. &~~ &~~      &~~        &~~     \\
\hline
  (1)                     &~~(2)            &~~~(3)        &~~(4)        &~~(5)       &~~~(6)          &~~~(7)    &~~~(8)    &~~(9)         &~~(10)  &~~~(11) &~~(12) \\
\hline
\hline
{\bf002616-351249}&~1.40$\pm$0.11    &~{\bf1.2}  &~30.5    &~1.31   ~&~~~0.07   &3.4e28 &~20.38   &1.996   &~G?/Bz?  &~FSRS     &~1  \\
 033427-015358    &~0.95$\pm$0.09    &~  --      &~  --    &~       ~&~~~       &       &~16.82   &1.487   &~G/Q   &~         &~AeB,2\\
 044245-681838    &~0.66$\pm$0.06    &~          &~37.6    &~.07    ~&~~~0.01   &       &~18.5    &0.964   &~G?   &~         &~1 \\
 084009-835432    &~0.76$\pm$0.18    &~1.4       &~ 4.8    &~.18    ~&~~~0.22   &3.6e26 &~18.13   &0.734  &~G?   &~FSRS &~2 \\
 094258-604621    &~0.59             &~ --       &   --    &~       ~&~~~       &       &~17.92   &0.586 ?  &~~~G   &~         &~  \\
 095744-153246    &~0.65$\pm$0.09    &~          &~14.5    &~.07    ~&~~~0.04   &       &~18.52   &1.787   &~~~G?   &~FSRS     &~AeB,2\\
 \hline\hline
 \end{tabular}
\label{tab:noHabbl}
\end{table*}

Sources that had reliable redshift measurements were classified in \cite{2011MNRAS.417.2651M} according to their spectral line characteristics into the following categories:
``Aa'' -- AGN with absorption lines only, ``Aae'' -- with absorption and emission lines, ``Ae'' -- with emission lines only, ``AeB'' -- with broad emission lines. 
The number ``1'' indicates that the source was classified in the SuperCOSMOS database \cite{2001MNRAS.326.1315H} as a galaxy, ``2'' -- as a stellar object showing a featureless spectrum in the continuum, especially in blue. 
Objects classified as Aa, Aae, Ae in the SuperCOSMOS database fall into the category of galaxies, and AeB are mostly associated with quasars. These data, given in columns (12), give us additional information about the objects belonging to one or another class.

All sources given in Table~\ref{tab:noHabbl} and having the largest deviations of $R$ from the Hubble relation, belong to objects that have no data at frequencies below 0.8 -- 1.4 GHz.
The source J002616-351249 is identified as a galaxy in the SuperCOSMOS database \cite{2001MNRAS.326.1315H}. However, as noted in \cite{2010MNRAS.408.1187H}, the strong variability of J002616-351249 at 95 GHz and at the frequencies of the WMAP mission (22\%, 38\% at 33 and 61 GHz, respectively) suggests that this is a "core-jet" source rather than a true HFP galaxy.

Sources J002616-351249, J084009-835432 and J095744-153246, according to the literature \cite{1995ASSL..203...95H} are flat-spectrum sources (FSRS), and they can hardly be classified as HFPs. Sources J033427-015358 and J095744-153246 have broad emission lines in the optical spectrum and belong to the ``AeB'' category, which is associated with quasars.
The sources J033427-015358, J084009-835432 and J095744-153246 in the SuperCOSMOS database are classified as stellar objects showing a featureless spectrum in the continuum, as indicated by the number 2 in column 12, and are most likely quasars or blazars.

For two sources in Table~\ref{tab:noHabbl}, J044245-681838 and J094258-604621, there is no evidence to challenge their classification as galaxies according to \cite{1995ASSL..203...95H}, but the question of whether they are genuine HFP galaxies remains open, since we have no information on their variability.

As for the sources with small deviations of $R$ values from the Hubble relation (Fig.~\ref{fig15}), and a list of which is given in Table~\ref{tab:yesHabbl}, among them there are also radio sources with a flat spectrum (J012346-092304, J033332-052301, J051321-212821,
J110957-373220), and they are most likely not genuine HFP galaxies.
In addition, the sources J051321-212821 and 112621-312358 were identified in the work of \cite{2011MNRAS.417.2651M} as ``AeB'', and may turn out to be quasars.

Thus, of the total number of sources initially identified with galaxies for which redshifts are known, only 27 can be considered candidates for genuine HFP galaxies and, possibly, young objects. 
However, even among them there are sources with uncertain identification (in the table they are marked with a question mark).

All 27 candidates have a peak frequency $\nu_{obs}$>5 GHz, and their radio luminosities at a frequency of 20 GHz lie in the range of $\sim 10^{23}-10^{26}$ W/Hz.
The sources J003033-581914, J015949-085000, J020835-173934 and J130031-441442 have data at the frequencies of the GLEAM and TGSS catalogs and therefore can be "restarted"\, HFP sources.

It is worth noting that for some of the sources identified as quasars, the $R--z$ dependence agrees well with the Hubble relation. This fact requires more thorough studies, primarily specifying the type of their parent objects. However, the classification into galaxies and quasars, which is based on optical morphology, is not always confident and reliable due to the weakness of the sample objects, and little information about their variability.
\begin{longtable*}{|l|l|l|l|l|l|l|l|l|l|l|l|}
\caption{\label{tab:yesHabbl}
Parameters of galaxies with minimal deviations of magnitudes $R$ from the Hubble relation: $\alpha_{below}$, FWHM, $\nu_{obs}$, $S_{peak}$), angular sizes $\theta$, radio luminosities $L_{\nu}$ at 20 GHz, redshifts $z$, magnitudes in the $R$ band, the type of radio source TypeR and the type of parent object Type, optical classification of objects from \cite{2011MNRAS.417.2651M}.
}
\\
\hline
\hline
 ~~NVSS      ~            &$\alpha_{below}$ &~FWHM         &~~$\nu_{obs}$  &~$S_{peak}$ &~~~~~$\theta$  &~~$L_{\nu}$   &~~~$R$     &~~$z$          &~Type  &~~TypeR   &~~opt.      \\
 ~~name      ~            &                 &~~            &~ GHz &~~Jy &~mas &~W/Hz &~mag. &~~ &~~      &~~        &~~     \\
\hline
  (1)                     &~~(2)            &~~~(3)        &~~(4)        &~~(5)       &~~~(6)          &~~~(7)    &~~~(8)    &~~(9)        &~~(10)  &~~~(11) &~~(12)   \\
\hline
\endfirsthead

\hline
  (1)                     &~~(2)            &~~~(3)        &~~(4)        &~~(5)       &~~~(6)          &~~~(7)   &~~~(8)    &~~(9)         &~~(10)  &~~~(11) &~~(12)    \\
\hline
\endhead

\hline
\endfoot

\hline\hline
\endlastfoot
 002705-050350    &~0.70$\pm$0.02 &~1.4 &~9.0  &~0.072 &~0.06 &~1.4e26 &~21.26 &~0.750 &~G &~ &~\\
 003033-581914*   &~1.02$\pm$0.30 &~    &~12.6 &~0.059 &~0.03 &~       &~16.23 &~0.174 &~G &~ &~Aa,1\\
 004905-552110    &~0.72$\pm$0.02 &~    &~22.0 &~0.090 &~0.02 &~       &~15.46 &~0.063 &~G &~ &~Aae,1\\
 011102-474911    &~0.63$\pm$0.02 &~1.3 &~27.1 &~0.090 &~0.02 &~5.9e24 &~17.38 &~0.154 &~G &~GPSc &~G,1\\
{\bf012346-092304}*&~0.90$\pm$0.27 &~{\bf1.3} &~5.5 &~0.180 &~0.17 &~1.0e25 &~17.01 &~0.143 &~G  &~FSRS &~G,1\\
 015949-085000*   &~0.66$\pm$0.14  &~          &~   &~      &~     &        &~17.57 &~0.405 &~G  &~ &~G,1\\
 020835-173934*   &~0.75$\pm$0.25  &~          &~10.7 &~0.240 &~0.08 &~      &~16.03 &~0.129   &~G &~&~1\\
 023611-420337    &~1.25           &~          &~15.9 &~0.100 &~0.04 &~      &~23.38 &~1.283   &~G? &~ &~ \\
 024710-632537    &~0.85$\pm$0.02  &~          &~16.6 &~0.090 &~0.03 &~      &~16.54 &~0.101   &~G &~ &~ \\
 025822-332705    &~0.83$\pm$0.09  &~          &~63.0 &~0.130 &~0.01 &~      &~24.39 &~1.216   &~G &~ &~ \\
 033332-052301    &~0.62$\pm$0.06  &~          &~15.9 &~0.080 &~0.03 &~       &~16.86 &~0.153 &~G  &~FSRS &~1\\
{\bf034941-540106}&~0.92$\pm$0.08  &~{\bf1.3} &~6.2   &~0.130 &~0.12 &~1.5e24 &~15.38 &~0.068 &~G &~ &~Aae,1\\
 040106-160640    &~0.64$\pm$0.19  &~          &~23.2 &~0.440 &~0.04 &~       &~13.80 &~0.032 &~G &~ &~Aae,1\\
 040438-041655    &~0.55$\pm$0.13  &~        &~    &~   &~     &~       &~22.73 &~0.687 &~G? &~ &~ \\
 042203-562127    &~0.58$\pm$0.13  &~1.5       &~9.0 &~0.100 &~0.06 &~3.5e23 &~14.93 &~0.038 &~G  &~ &~Aa,1\\
 051321-212821*   &~0.92$\pm$0.30  &~        &~    &~     &~       &        &~17.28 &~0.356   &~G &~FSRS &~AeB,1\\
 054828-331331    &~0.87$\pm$0.11  &~          &~33.6 &~0.060 &~0.01 &~      &~15.09 &~0.041  &~G &~ &~Aa,1\\
{\bf063004-551751}&~1.28           &~{\bf1.3} &~ 6.7 &~0.080 &~0.09 &~6.8e23 &~14.14 &~0.058 &~G  &~ &~Aae,1\\
 070949-381152    &~1.80           &~          &~20.6 &~0.090 &~0.02 &~       &~17.75 &~0.126   &~G  &~ &~Aa,1\\
 074109-544746    &~0.63$\pm$0.15 &~       &~     &~     &~       &~       &~17.95 &~0.106 &~G &~ &~Aae,1\\
 083821-071336    &~0.61$\pm$0.08 &~        &~    &~  &~    &~       &~16.47 &~0.117 &~G &~ &~1\\
{\bf091900-253350}&~1.83$\pm$0.07 &~{\bf1.2} &~ 7.0 &~0.110 &~0.09 &~6.6e24 &~18.44 &~0.148 &~G  &~ &~   \\
 110957-373220*   &~0.84           &~          &~48.6   &~0.080  &~~~0.01   &~       &~11.11 &~0.010  &~~~G  &~FSRS     &~Aa,1\\
 112621-312358    &~0.94$\pm$0.17  &~          &~11.0   &~0.070  &~~~0.05   &~       &~17.88 &~0.321 &~~~G  &~         &~AeB,1\\
 114503-325824    &~0.61$\pm$0.09  &~1.5       &~13.6   &~0.080  &~~~0.03   &~2.8e23 &~13.76 &~0.038 &~~~G  &~         &~Aae,1\\
 123449-243232    &~0.66$\pm$0.03  &~1.4       &~10.5   &~0.090  &~~~0.05   &~8.1e24 &~16.17 &~0.181  &~~~G  &~         &~Aa,1\\
 130031-441442*   &~0.77$\pm$0.09  &~1.4       &~ 7.1   &~.11  &~~~0.09   &~2.7e23 &~9.21  &~0.032   &~~~G  &~         &~G,1\\
{\bf181225-712006}&~0.69           &~{\bf1.3}  &~13.0   &~0.040  &~~~0.03   &~4.5e24 &~16.84 &~0.199   &~~~G  &~         &~Aa,1\\
 194131-760548    &~0.71$\pm$0.02  &~        &~    &~  &~~~     &~     &~16.53 &~0.146   &~~~G  &~         &~Aae,1\\
 210925-361557    &~0.92$\pm$0.08  &~          &~14.6   &~0.070  &~~~0.03   &~       &~18.28 &~0.226   &~~~G  &~         &~Aa,1\\
 220413-465424    &~1.96           &~          &~13.3   &~0.120  &~~~0.05   &~       &~21.8  &~0.832   &~~~G? &~         &~ \\
 225558-103922    &~0.80$\pm$0.10  &~          &~14.1   &~0.100  &~~~0.04   &~       &~20.74 &~0.616   &~~~G? &~         &~ \\
 231546-230744    &~0.93$\pm$0.17  &~          &~12.5   &~0.080  &~~~0.04   &~       &~17.05 &~0.069   &~~~G  &~         &~Aae,1\\
\hline\hline
\end{longtable*}
\section{Variability of radio emission of sources.}

For quantitative estimates of the variability of the flux densities of sources, the variability index $Var_{S}$ was used, which is calculated using the formula \cite{1992ApJ...399...16A}: \\
$Var_{S}$=$\frac{(S_{max}-\sigma_{max})-(S_{min}+\sigma_{min})}{(S_{max}-\sigma_{max})+(S_{min}+\sigma_{min})}$, (4) \\
where $S_{max}$ and $S_{min}$ are the maximum and minimum values of the flux densities, $\sigma_{max}$ and $\sigma_{min}$ are errors of the flux densities.

Weakly variable sources with $Var_{S}$ not exceeding 25\% \citep{1992ApJ...399...16A} are considered candidates for HFP.

The variability index was calculated only for radio sources for which the CATS database contained data on flux densities obtained from observations with one telescope and at least at one frequency over a certain period of time, as well as data on measurement errors.

The list of sources that met these requirements and whose variability indices on scales of several years were greater than 10\% is given in Table~\ref{tab:Var}. The table contains the value of the variability index $Var_{S}$, the frequency $\nu_{var}$ at which the variability index was calculated, the peak frequency $\nu_{obs}$, and the name of the catalog from which the flux densities were taken. 
Asterisks mark radio sources that have data in the TGSS and GLEAM catalogues.
\begin{table}
\caption{
Sources of the sample  with variability indices $Var_{S}$>10\%.\\ Variability index $Var_{S}$, the frequency $\nu_{var}$ at which the variability index was calculated, the peak frequency $\nu_{obs}$, the name of the catalog from which the flux densities. $Cat.$ -- symbols marks catalogs and references: (a) -- ATPMN~\citep{2012MNRAS.422.1527M}, (b) -- GPS2~\citep{Tr}, (c) -- VLASS~\citep{2020PASP..132c5001L}, (d) -- CGR15~\citep{2011ApJS..194...29R}, (e) -- AT95G~\citep{2008MNRAS.385.1656S}, (f) -- SEST3~\citep{1996A&AS..116..157T}, (g) -- ATC18~\citep{2004MNRAS.354..305R}.
}
\begin{tabular}{|l|c|c|c|l|}
\hline\hline
 ~~NVSS       &~$Var_{S}$ &~$\nu_{var}$ &~$\nu_{obs}$ &~Cat. \\
 ~~name       &~~~\%      &~GHz         &~GHz         &~~          \\
\hline
~~~~~(1)       &~(2)     &~~~(3)         &~~(4)        &~~(5) \\
\hline
~012407-730904  &~14  &~4.8    ~&~6.5  &~(a) \\
~034941-540106  &~10  &~4.8    ~&~6.2  &~(a) \\
~042203-562127  &~22  &~4.8    ~&~9.0  &~(a) \\
~050555-293038*~&~30  &~21.0   ~&~6.0  &~(b) \\
~080633-291135 ~&~17  &~4.8    ~&~5.0  &~(b) \\
~094219-231703*~&~48  &~3.0    ~&~      &~(c) \\
~095727-015655 ~&~25  &~15.0    ~&~6.8  &~(d) \\
~110957-373220*~&~17  &~20.0    ~&~48.6  &~(e) \\
~111719-483809*~&~11  &~20.0    ~&~29.2  &~(e) \\
~112953-024006 ~&~22  &~15.0    ~&~9.2  &~(d) \\
~113143-581853 ~&~19  &~4.8    ~&~9.2  &~(a) \\
~142741-330531*~&~11  &~20.0    ~&~89.8  &~(e) \\
~153851-165526 ~&~27  &~15.0    ~&~8.9  &~(d) \\
~200324-042137 ~&~10  &~15.0    ~&~6.1  &~(d) \\
~201115-154640*~&~27  &~90.0    ~&~20.8  &~(f) \\
~201500-671258 ~&~12  &~18.5   ~&~15.3  &~(g) \\
~205625-320845*~&~98  &~3.0    ~&~39.1  &~(c) \\
~212402-602808*~&~15  &~20.0    ~&~12.5  &~(e) \\
~223015-132543*~&~16  &~15.0    ~&~14.1  &~(d) \\
~230737-354828 ~&~12  &~20.0    ~&~19.8  &~(e) \\
~235311-274324*~&~76  &~3.0    ~&~16.6  &~(c) \\
\hline\hline
\end{tabular}
\label{tab:Var}
\end{table}
Based on the information provided by the CATS database, only 21 sources had a variability index $Var_{S}$ of $\ge$ 10\%.
Variability of 25\% or higher was found in seven radio sources: J050555-293038, J094219-231703, J095727-015655, J153851-165526, J201115-154640, J205625-320845, and J235311-274324.
All of them are quasars; two of them, J094219-231703 and J201115-154640, are blazars.

Another source, J002616-351249, is considered variable based on the data at 95 GHz (AT95G) \cite{2010MNRAS.408.1187H} and at the frequencies of the WMAP mission \cite{2003ApJS..148....1B, 2009ApJS..180..283W, 2009MNRAS.392..733M}: 22\%, 38\% at 33 and 61 GHz, respectively.
The variability index of the source J002616-351249 at 95 GHz, calculated from the data from \cite{2010MNRAS.408.1187H} using formula (4), was 37\%.

Note that half of the radio sources presented in the table have data in the TGSS and GLEAM catalogs, i.e. they can belong to radio sources that exhibit non-thermal emission of lobes on scales from kpc to mpc.

Sources J050555-293038, J095727-015655, J153851-165526 and J201115-154640 show variability at frequencies above the peak frequency, suggesting that we are observing jet emission and that these sources are not "genuine" HFPs.

\section{CONCLUSION}

To study the spectral properties of radio sources with a high-frequency peak in the spectrum, a sample of sources from the AT20G catalog was formed, the spectral indices of which, $\alpha_{below}$, characterizing the radiation in an optically thick medium, exceed +0.5. 
The sample includes 269 sources, most of which $\sim$ 70\% are quasars, of which $\sim$ 25\% are blazars. The number of galaxies is 2.5 times less than quasars.

Using the source flux density data from the CATS database, spectra were constructed for all objects in the sample and their main parameters were determined:
spectral indices of radiation in optically thick and thin media $\alpha_{below}$ and $\alpha_{above}$, respectively, peak frequencies in the observer's reference frame $\nu_{obs}$ and in the source's reference frame $\nu_{int}$, flux densities at the peak frequency $S_{peak}$ and the spectral half-width FWHM.
Note that when constructing spectra, we often use heterogeneous data obtained at different telescopes and in different observation epochs.

Almost all sources in the sample are HFP if only the limiting peak frequency $\nu_{obs}\geq$5 GHz is used as the HFP criterion. Most sources have flux densities below 0.3 Jy. 43 sources have inverted spectra in the range of frequencies under consideration.

About 70\% of the radio sources in the sample are sources that do not have flux density data in the low-frequency region of the spectrum, namely at the frequencies of the GLEAM and TGSS catalogs. The average $\alpha_{below}$ values of the sources in this group are 30\% higher than those of the sources that have flux density data at frequencies below 0.8 - 1.4 GHz, and the average values of peak flux densities and peak frequencies $\nu_{obs}$ are 24\% and 37\% lower, respectively.

A comparison of spectral parameters of sources identified with galaxies and quasars was performed. Distributions of $\alpha_{below}$ of these populations have almost identical maxima: 0.70$\pm$0.03 for quasars and 0.73$\pm$0.02 for galaxies and the FWHM half-width. At the same time, the average values of $\alpha_{below}$ of galaxies are approximately 8\% higher than those of quasars. 
In the article \cite{2013AstBu..68..262M} this difference was $\sim $ 10\%.

The average values of peak frequencies $\nu_{int}$ of quasars and galaxies are 40.4 GHz and 28.9 GHz, respectively, the median values of $z$ are 0.15 and 1.0, respectively, and the peak flux densities differ significantly. Most galaxies have peak flux densities below 0.2 Jy, while for quasars they are below 0.4 Jy. The average $S_{peak}$ values are 0.13$\pm0.17$ and 0.24$\pm0.34$, respectively.
It was previously expected by \cite{2010MNRAS.408.1075O} that young objects would be more common among galaxies with flux densities below 0.3 Jy.
This increases the likelihood that young HFP sources may be present among the galaxies in our sample.
The vast majority of galaxies are radio sources with no data at frequencies below 0.8 GHz.

We compared the characteristics of galaxies and quasars in our sample with the results obtained in \cite{2009AN....330..180H}, which analyzes a sample of 656 GPS sources from the AT20G catalog, which was formed based on the constraint on the peak frequency of $\nu_{obs}$>5 GHz.

The comparison showed that the proportion of quasars and galaxies in the samples almost coincided, and the average redshifts of galaxies and quasars in the samples were also close.

At the same time, the average flux densities of galaxies and quasars in our sample are almost two times smaller than in \cite{2009AN....330..180H}.
In addition, the peak frequencies of the sources in the \cite{2009AN....330..180H} sample are in the range of 8–15 GHz, while for most sources in our sample the peak frequency range $\nu_{obs}$ is 5–50 GHz.

Among the radio sources in our sample, 187 sources with ultra-inverted spectra were found, for which $\alpha_{below} \geq +0.7$. Such sources are the most likely HFP candidates, and their spectra may indicate the youth of the radio source \cite{2010MNRAS.405.1560M}.
Sources with ultra-inverted spectra accounted for 3.2\% of all sources in the AT20G catalog and 70\% of all sources in our sample.
42 of them have data at the frequencies of the TGSS and GLEAM catalogs.

The radio luminosities $L_{20}$ at a frequency of 20 GHz, the angular sizes $\theta$ of the emitting regions, and the linear sizes LS of the radio sources in the sample were estimated. The ranges of radio luminosities of the galaxies and quasars in the sample at a frequency of 20 GHz coincided with those obtained in \cite{2011MNRAS.417.2651M}, as did the nature of the dependence of $L_{\nu}$ on $\nu$. For galaxies, the range of radio luminosities was $10^{23}-10^{26}$ W/Hz, for quasars — $10^{26}-10^{30}$ W/Hz.

The values of $\theta$ are in the range of 0.002 -- 0.25 mas and are significantly smaller than in \cite{1990A&A...231..333F, 1990cssg.conf...55S, 2019AstBu..74..348S}. The linear sizes of LS are in the range from 0.2 pc to 30 pc. For galaxies, LS < 30 pc, for most quasars < 10 pc. When estimating LS, an analytical expression was used to relate the peak frequency $\nu_{int}$ and the largest linear size of sources from \cite{2014MNRAS.438..463O}.

The greatest difficulties arose with estimating the variability of the sources.
The spectra of many sources are uninformative. For example, in the optically thin part of the spectrum, data are missing or very sparse. Although the sample contains many blazars and flat-spectrum sources, many of them do not have flux density measurement errors in the CATS database. Thus, the variability index could be estimated only for a very limited number of objects: those for which the CATS database had flux density and root-mean-square error data measured at one telescope over a certain period of time and at least at one frequency. For 21 sources, the variability index $Var_{S}$ was $\ge$ 10\%, and for seven quasars, two of which are blazars, it was more than 25\%.

Difficulties with variability estimates also affected the possibility of identifying sources in the sample that can be considered "genuine" HFPs. According to the criteria in \cite{2010MNRAS.408.1187H}, ``genuine'' HFPs, in addition to the $\nu_{peak}$ > 5 GHz condition, should have weak flux density variability, compact radio morphology, and their optical spectrum should differ from that of blazars. In our sample, which mainly consists of quasars (blazars) and flat-spectrum sources, ``genuine'' HFPs are likely to be very few and difficult to detect. Additional information on their optical properties, morphology, and variability on long time scales will be needed. Most likely, they should be sought among galaxies, since they have lower variability.

For the galaxies in the sample, the dependences of the stellar magnitude in the $R$ band on the redshift were constructed. Comparing this dependence with the Hubble relation found for the GPS in \cite{1996MNRAS.279.1294S}, 27 galaxies were identified for which the deviations of $R$ from this relation are small. These galaxies can be considered candidates for genuine HFPs and, possibly, candidates for young radio sources.
Looking through the radio spectra of the sources in the sample, we found approximately the same number of objects ($\sim$ 25) that do not have obvious signs of variability and there is no reliable data that they are FSRS. These are mainly galaxies. However, it is too early to claim that they are "genuine"\, HFPs.
Further research is required.

\begin{acknowledgments}
The work was carried out within the framework of the state assignment of SAO RAS, approved by the Ministry of Science and Higher Education of the Russian Federation.
VizieR.
This research has made use of the NASA/IPAC Extragalactic Database, which is funded by the National Aeronautics and Space Administration and operated by the California Institute of Technology.
When constructing radio spectra, the database of radio astronomy catalogs CATS \citep{2005BSAO...58..118V} was used,
and the FADPS system for processing radio astronomy data \citep{1993BSAO...36..132V, 1997ASPC..125...46V} was also used.
\end{acknowledgments}

\section*{CONFLICT OF INTEREST}
The authors declare no conflicts of interest.

\bibliographystyle{aspb1}
\bibliography{HPF}

\clearpage
\newpage

\appendix
\renewcommand{\thesection}{\Alph{section}.\arabic{section}}
\setcounter{section}{0}

\section{The sample of HFP sources}
The appendix contains a table with the parameters of the HFP sources from the prepared sample. It also contains a description of the table columns and examples of the most characteristic spectra of some radio sources.
\subsection{Table} \label{Tab}
The columns contain the following information: \\
\begin{enumerate}
\item[(1)] $NVSS$ is an source name from NVSS catalog \citep{1998AJ....115.1693C}. The asterisk ``*'' marks the sources that have low-frequency flux density data from the GLEAM \citep{2017MNRAS.464.1146H} and TGSS \citep{2017A&A...598A..78I} catalogs. In this part of the spectrum, such sources have a negative spectral index, which indicates the presence of extended non-thermal radio emission of lobes on scales from tens of kiloparsecs to megaparsecs.
Such sources may include restarted HFP radio sources. Radio sources whose spectra satisfy the condition: $\alpha_{below}\ge +0.7$ and FWHM $\leq$1.35 are highlighted in bold; 
\item[(2)] spectral index below the peak frequency in the spectrum $\alpha_{below}$;
\item[(3)] spectral index above the peak frequency $\alpha_{above}$;
\item[(4)] The width of the spectral peak at half power at the peak frequency $FWHM$ in frequency decades. The half-widths of the peaks of sources whose spectra satisfy the condition: $\alpha_{below}\ge +0.7$ and FWHM $\leq$1.35, are shown in bold;
\item[(5)]  peak frequency in the rest frame of the observer $\nu_{obs}$ in GHz;
\item[(6)]peak frequency in the rest frame of the source $\nu_{int}$ in GHz;
\item[(7)]spectral flux density at the peak frequency $S_{peak}$ in Jy;
\item[(8)]estimated angular size of the emission region $\theta$ in milliarcseconds (mas);
\item[(9)]radio luminosity at a frequency of 20 GHz $L_{20}$ in W/Hz;
\item[(10)]redshift $z$. Redshift information was taken from the NASA Extragalactic (NED) \cite{1995ASSL..203...95H}, Simbad \cite{2000A&AS..143....9W}, NOIR DataLab \cite{2019BAAS...51g..61O,2019arXiv190800664O}, SDSS and Vizier \cite{2000A&AS..143...23O} databases. The symbol ``s'' denotes the spectroscopic redshift, ``p'' -- the photometric redshift;
\item[(11)] $Type$  combines the type of the parent object and the type of the radio source. Optical types of parent objects are presented as ``G'' — galaxy, ``Q'' — quasar, ``Bz'' — blazar, ``BL'' — BL Lac stars, ``PNe'' — planetary nebula, ``HII'' — hydrogen ionized region, ``S'' — star, ``YSO'' — young stellar object. To determine the type of the parent object, the Simbad and NED databases were used. The types of radio sources are presented as FSRQ — flat-spectrum radio quasar (-0.5 < $\alpha$ < 0, $S\sim e^{\alpha}$) according to the \citep{2009A&A...495..691M} classification, FSRS — flat-spectrum radio source.
The "?" sign next to the optical or radio type indicates that the type is not determined with certainty;
\item[(1)] Comments $Com.$. Explanations for this column:
\begin{itemize}
\item ``at'' -- flux density data are available only at AT20G frequencies (4.8, 8.6, and 20 GHz), ``3f'' -- data are available only at 3 frequencies;
\item ``PL'' -- data are available for the radio source in the Planck catalog \citep{2014A&A...571A..28P};
\item ``t8'', ``t9'' and ``t89'' mark radio sources that do not have data in the NVSS or SUMSS \citep{1999AJ....117.1578B,2003MNRAS.342.1117M} catalogs (see \cite{2010MNRAS.402.2403M}, table 8) and/or whose spectral index is $\alpha^{5}_{20} \ge +0.7$ (see \cite{2010MNRAS.402.2403M}, table 9 );
\item ``SUM'', ``VLASS'',``WISH'' — the ``+'' symbol marks sources for which data from SUMSS, VLASS \citep{2021ApJS..255...30G}, WISH \citep{2002A&A...394...59D} were used to estimate spectral parameters, the ``--'' symbol indicates that these data were not used to construct spectra;
\item $TGSS$ — in the low-frequency part of the radio source spectrum there are only data from the TGSS catalog.
\end{itemize}

\begin{longtable*}{|l|l|l|c|c|c|c|c|c|l|l|l|}
\caption{\label{tab:Sp.45+}
Parameters of the spectra of radio sources in the sample: $\alpha_{below}$ and $\alpha_{above}$ are the spectral index below and above the peak in the spectrum, $FWHM$ is the half-width of the spectral peak, $\nu_{obs}$ is the observed frequency of the peak, $\nu_{int}$ is the frequency of the peak in the rest frame of the radio source, $S_{peak}$ is the flux density at the frequency of the peak, $\theta$ is the estimate of the angular size, $L_{20}$ is the radio luminosity at 20 GHz, $z$ is the redshift, $Type$ is the host type and the type of the radio source, $Com.$ are comments.
}
\\
\hline\hline
~NVSS      ~        & $\alpha_{below}$ &~~$\alpha_{above}$ & FWHM  &~$\nu_{obs}$  &~$\nu_{int}$   &~ $S_{peak}$  & $\theta$       &~$L_{20}$    &~$z$   & $Type$  &~$Com.$          \\
	     ~                &             &              &~              &~GHz &~GHz &~Jy      & mas &~W/Hz &~~     &~~                &~~       \\
\hline
~~~~(1)              &~~~~(2)      &~~~~~(3)     &~~~(4)      &~~(5)     &~(6)      &~~(7)     &~ (8)    &~~~(9)    &~ (10)  & (11)  &~~(12)      \\
\hline
\endfirsthead

\hline
  ~~~~(1)              &~ (2)      & (3)          &   (4)      &  (5)     & (6)      &  (7)     &  (8)    & (9)    & (10)  & (11)  & (12)    \\
\hline
\endhead

\hline
\endfoot

\hline\hline
\endlastfoot
~001948-173446* &~0.60$\pm$0.04 &               &~~~          &~98.0  &~       &~0.22   &~      &~           ~&~                &~Q  &    \\
 002343-593032    &~0.80$\pm$0.11 &~-0.20         & 1.4     &~ 8.6  &~14.4  &~0.11  &~      &~1.6e26 ~&~0.677,p &~Q  &~   \\
 002442-420203    &~0.97$\pm$0.12 & -1.19$\pm$0.07 &~1.4     &~ 2.0  &~~3.9  &~2.80   &~      &~1.5e28 ~&~0.937,s &~Q &~   \\
 {\bf002514-094200} &~0.99$\pm$0.10 &-0.44        &  {\bf1.3} & 8.9 &      & 0.08 &~      &~       ~&~        &~G? &~    \\
 {\bf002616-351249} &~1.40$\pm$0.11 & -0.85$\pm$0.03&~{\bf1.2} &~20.1 &~60.2 &~0.80 &~0.07  &~3.4e28 ~&~1.996,s &~G/Bz? &~t9  \\
 &~ & &~ & &~ &~ &~  &~ ~&~ &~FSRS &~ \\
  002628-740020    &~0.92$\pm$0.04 &            &~        &~16.0  &~      &~0.09 &~      &~       ~&~        &~Q? &~    \\
  002705-050350    &~0.70$\pm$0.02 & -0.48       &~1.4    &~9.0 &~15.8 &~0.07 &~0.06 &~1.4e26 ~&~0.750,p &~G &~    \\
 003033-581914*   &~1.02$\pm$0.30 &               &~          &~12.6 &~14.8 &~0.06 &~0.03 &~ &~0.174,s &~G  &~3f  \\
 003207-154132 &~0.69$\pm$0.05 &               &~         &~15.6 &~25.5 &~0.11 &~   &~ &~0.635,s &~Q  &~--WISH\\
  &~ &    &~   &~ &~ &~ &~   &~ &~ &~FSRS &~\\
 003816-012204 &~0.94$\pm$0.16 &               &~          &~10.8 &~23.9 &~0.12 &~      &~       ~&~1.215,s &~Q &~    \\
 004417-375259* &~0.69$\pm$0.06 &               &~          &~34.6 &~51.4 &~0.11 &~      &~       ~&~0.483,s &~Q/Bz &     \\
  &~ &    &~   &~ &~ &~ &~      &~       ~&~ &~FSRQ &     \\
 004905-552110 &~0.72$\pm$0.02 &               &~         &~22.0 &~23.4 &~0.09 &~0.02 &~       ~&~0.063,s &~G &~    \\
 004937-705211 &~0.76$\pm$0.07   &               &~         &~   &~      &~      &~      &~       ~&~       &~Q? &~    \\
 005920-612647* &~0.74$\pm$0.004  &               &~          &~11.8 &~27.1 &~0.08 &~      &~       ~&~1.294,p &~Q &     \\
 011102-474911 &~0.63$\pm$0.02 & -0.50       &~1.3 &~27.1 &~31.3 &~0.09 &~0.02 &~5.9e24 ~&~0.154,s &~G       &~t9,--SUM\\
               &~0.49$\pm$0.07 &             &~1.6 &~31.4 &~36.2 &~0.07 &~      &~       ~&~      &  &~+SUM \\
{\bf012346-092304}* &~0.90$\pm$0.27 & -0.59 &~{\bf1.3} &~5.5 &~6.2 &~0.18 &~0.17  &~1.0e25 &~0.143,s &~G  &~  \\
 &~ &  &~ &~ &~ &~ &~  &~ &~ &~FSRS? &~  \\
 012407-730904 &~0.56$\pm$0.05 & -0.09$\pm$0.06 &~1.6 &~6.5 &~6.5 &~0.11 &~      ~&~7.3e19 ~&~0.0005,s       &~HII         &~PL \\
  &~ &  &~ &~ &~ &~ &~      ~&~ ~&~ &~FSRS     &~ \\
{\bf012744-345755} &~1.16$\pm$0.01 & -0.54$\pm$0.14 &~{\bf1.3} &~7.8 &~24.7 &~0.08 &~      &~1.5e27 ~&~2.163,s &~Q      &~--SUM\\
 &~ & &~ &~ &~ &~ &~      &~ ~&~ &~FSRS     &~--SUM\\
                 &~0.79$\pm$0.07 &-0.54$\pm$0.14 &~1.4 &~9.9 &~31.2 &~0.07 &~      &~1.3e27~&~ &~ &~+SUM\\
 013210-795905 &~0.50$\pm$0.002 & -0.54$\pm$0.16 &~1.5 &~5.0 &~10.4 &~0.07 &~      &~3.3e26 &~1.077,p &~Q          &~   \\
  &~ & &~ &~ &~ &~ &~      &~ &~ &~FSRS     &~   \\
 013658-133534* &~0.51$\pm$0.09 &               &~         &~  --  &~      &~      &~      &~       ~&~1.126,p &~Q           &~   \\
 013707-244447    &~0.85$\pm$0.07 &-0.42$\pm$0.12 &~1.4       &~11.9 &~24.5 &~0.09 &~      &~3.8e26 ~&~1.050,s &~Q              &~   \\
 014207-420601* &~0.49$\pm$0.04 & -0.46$\pm$0.13 &~1.6 &~18.3 &~33.8  &~0.10 &~    &~2.6e26 ~&~0.842,p &~Q?              &~  \\
 014225-572957* &~0.71          &               &             &~13.5 &~23.9 &~0.08 &~      &~       ~&~0.773,p &~Q             &~at\\
 014310-320055* &~0.50$\pm$0.15 & -0.48$\pm$0.17 &~~1.5       &~36.2  &~49.8  &~0.51   &~      &~2.3e26 ~&~0.375,s &~Q/Bz       &~  \\
 &~ & &~ &~ &~ &~ &~      &~ &~ &~FSRS     &~   \\
 015719-382944* &~0.73$\pm$0.12 &-0.51$\pm$0.09 &~~1.5       &~23.1 &~43.5  &~0.16   &~      &~4.9e26 ~&~0.888,p &~Q/Bz?           &~  \\
 015913-470614    &~0.76$\pm$0.15 & -0.33$\pm$0.06 &~~1.4       &~ 8.3  &~14.2  &~0.08   &~      &~1.4e26 ~&~0.713,p &~Q           &~   \\
 &~ & &~ &~ &~ &~ &~      &~ &~ &~FSRS     &~   \\
 015949-085000*   &~0.66$\pm$0.14   &-              &~~~          &~  &~      &~      &~      &~       ~&~0.405,s &~G              &   \\
{\bf020051-154237}&~0.95$\pm$0.11   & -0.79$\pm$0.23 &~~{\bf1.3}  &~ 7.5  &~15.0  &~0.17 &~      &~8.4e26 ~&~1.014,p &~Q/Bz?     &~   \\
&~ & &~ &~ &~ &~ &~      &~ &~ &~FSRS     &~   \\
 020835-173934*   &~0.75$\pm$0.25   &               &~~          &~10.7  &~12.1  &~0.24   &~0.08  &~       ~&~0.129,s &~G               &~  \\
 021229-255818    &~1.13$\pm$0.14   &               &~~~          &~11.8  &~      &~0.07   &~      &~       ~&~        &~G?              &~   \\
 023111-474612* &~0.72$\pm$0.21   &-0.53$\pm$0.04 &~~~1.5       &~29.5  &~52.1  &~0.46   &~      &~1.0e27 ~&~0.765,s &~Q/Bz    &   \\
 023611-420337    &~1.25            &               &~~~          &~15.9  &~36.4  &~0.10   &~0.04  &~       ~&~1.283,p &~G?              &~at,\,t8\\
 024709-281049*   &~0.70$\pm$0.09   &               &~~~          &~29.0  &~52.3  &~0.14   &~      &~       ~&~0.803,p &~Q               &~t9 \\
 024710-632537    &~0.85$\pm$0.02   &               &~~~          &~16.6  &~18.3  &~0.09   &~0.03  &~       ~&~0.101,p &~G               &~   \\
 025055-361635    &~1.48$\pm$0.15   &               &~~~          &~25.5  &~43.9  &~0.32   &~      &~       ~&~0.725,s &~Q/Bz?           &~t9 \\
 025822-332705    &~0.86$\pm$0.14   &               &~~~          &~19.4  &~43.0  &~0.10   &~0.03  &~       ~&~1.216,p &~G            &~--SUM\\
                 &~0.83$\pm$0.09   &               &~~~          &~63.0  &139.6  &~0.13   &~      &~       ~&~           &~         &~+SUM\\
 025928-001959*   &~0.68$\pm$0.04   & -0.62          &~~1.5       &~19.4 &~58.2 &~0.87 &~      &~1.7e28 ~&~2.000,s &~Q/Bz            &   \\
 030036-704448    &~0.74$\pm$0.03   &-0.47          &~~1.4       &~ 5.7  &~21.2  &~0.09   &~      &~2.9e27 ~&~2.725,p &~Q         &~   \\
 &~ & &~ &~ &~ &~ &~      &~ &~ &~FSRS     &~   \\
 030406-450342    &~1.08            &               &~~~          &~15.7  &~31.4  &~0.06   &~      &~       ~&~1.000,s &~Q              &~at,\,t8 \\
 030457-603555    &~1.10$\pm$0.02   &               &~~~          &~17.9  &~40.6  &~0.10   &~      &~       ~&~1.269,p &~Q              &~   \\
 032213-345832*   &~0.48$\pm$0.11   &               &~~~          &~   &~      &~      &~      &~       ~&~0.950,p &~Q             &~  \\
 032743-180342*   &~0.79$\pm$0.08   &               &~~~          &~13.2  &~26.6  &~0.23   &~      &~       ~&~1.015,p &~Q/Bz?           &~  \\
 033332-052301    &~0.62$\pm$0.06   &               &~~~          &~15.9  &~18.3  &~0.08   &~0.03  &~       ~&~0.153,p &~G/BL        &~   \\
 &~ & &~ &~ &~ &~ &~      &~ &~ &~FSRS     &~   \\
 033427-015358    &~0.95$\pm$0.09   &               &~~~          &~   &~      &~      &~      &~       ~&~1.480,s &~G               &~   \\
 033540-311901*   &~0.71$\pm$0.02   &               &~~~          &~14.5  &~24.3  &~0.17   &~      &~       ~&~0.680,p &~Q               &~  \\
{\bf034258-431813}&~1.55            &               &~~~          &~12.4  &~13.8  &~0.13   &~      &~       ~&~0.119,p &~Q               &~3f,\,t8 \\
 034650-771437    &~0.63$\pm$0.19   &               &~~~          &~19.8  &~      &~0.05   &~      &~       ~&~       &~G/BL?          &~+SUM\\
{\bf034941-540106}&~0.92$\pm$0.08   &-0.31          &~{\bf1.3}  &~ 6.2  &~~6.6  &~0.13   &~0.12  &~1.5e24 ~&~0.068,s &~G               &~   \\
 040019-225624*   &~0.50$\pm$0.02   &               &~        &~   &~      &~      &~      &~       ~&~        &~Q/Bz?       &~   \\
 &~ & &~ &~ &~ &~ &~      &~ &~ &~FSRS     &~   \\
	        &~1.03$\pm$0.05 &               &~{\bf1.3}  &~14.2  &~      &~0.25   &~      &~       &~  &~           &~   \\
 040039-575712    &~0.60$\pm$0.14   &-0.34          &~1.5       &~ 7.8  &~18.2  &~0.07   &~      &~4.6e26 ~&~1.349,p &~Q               &~   \\
 040106-160640    &~0.64$\pm$0.19   &               &~          &~23.2  &~23.9  &~0.44   &~0.04  &~       ~&~0.032,s &~G/Bz           &~   \\
 040438-041655    &~0.55$\pm$0.13   &               &~         &~  --  &~      &~      &~      &~       ~&~0.687,p &~G?              &~   \\
 040446-294012*   &~1.52$\pm$0.22   &               &~~~          &~11.4  &~26.4  &~0.08   &~      &~       ~&~1.317,p &~Q              &~--VLAS\\
	         &~0.68$\pm$0.06  &               &~         &~16.8 &~38.9 &~0.06   &~    &~  ~&~~      &~         &~+VLAS\\
 041514-402341*   &~0.80$\pm$0.07   &               &~          &~23.3  &~55.6  &~0.12   &~      &~       ~&~1.389,s &~Q/Bz            &~   \\
 041959-563611    &~0.51            &-0.57          &~~~1.5       &~ 4.5  &~11.5  &~0.07   &~      &~7.4e26 ~&~1.529,p &~Q/Bz?      &~   \\
 &~ & &~ &~ &~ &~ &~      &~ &~ &~FSRS     &~   \\
 042119-672902    &~0.78            &               &~~~          &~12.7  &~28.6  &~0.14   &~      &~       ~&~1.253,p &~Q?             &~--SUM\\
               &~0.40$\pm$0.05   &               &~~~          &~59.1  &133.2  &~0.14   &~      &~       ~&~     &~             &~+SUM\\
 042203-562127    &~0.58$\pm$0.13   &-0.13          &~~~1.5       &~ 9.0  &~~9.3  &~0.10   &~0.06  &~3.5e23 ~&~0.038,s &~G               &~   \\
 042810-435746    &~0.87$\pm$0.12   &-0.50          &~~~1.4       &~18.7  &~46.2  &~0.10   &~      &~9.1e26 ~&~1.472,p &~Q?             &~t9 \\
 042810-643823*   &~0.51$\pm$0.09   &-0.51          &~~~1.4       &~23.7  &~      &~0.34   &~      &~       ~&~       &~Q/Bz?          &~   \\
{\bf043121-575005}&~1.34$\pm$0.02   &-0.52          &~~~{\bf1.3}  &~ 7.5  &~      &~0.11   &~      &~       ~&~        &~G?         &~   \\
&~ & &~ &~ &~ &~ &~      &~ &~ &~FSRS     &~   \\
{\bf043229-161405}*&~0.94$\pm$0.18  &-0.38          &~~~{\bf1.3}  &~ 7.8  &~      &~0.35   &~      &~       ~&         &~Bz?            &~   \\
{\bf043429-234205}*&~0.91$\pm$0.23  &-0.63$\pm$0.19 &~~~{\bf1.3}  &~ 5.8  &~11.5  &~0.15   &~      &~6.1e26 ~&~0.979,s &~Q/BL       &~TGSS \\
&~ & &~ &~ &~ &~ &~      &~ &~ &~FSRS     &~   \\
{\bf044023-473218}&~1.39$\pm$0.02   &               &~~~{\bf1.3}  &~37.3  &~74.8  &~0.07   &~      &~       ~&~1.004,p &~Q?             &~t89\\
 044047-695217    &~0.64$\pm$0.06   &-0.48$\pm$0.03 &~~~1.4       &~37.7  &112.7  &~0.92   &~      &~1.5e28 ~&~1.989,p &~Q          &~   \\
 &~ & &~ &~ &~ &~ &~      &~ &~ &~~FSRS     &~   \\
 044245-681838    &~0.66$\pm$0.06   &               &~~~          &~37.6  &~73.8  &~0.07   &~0.01  &~       ~&~0.964,s &~G               &~   \\
{\bf044854-291612}&~0.87$\pm$0.02   &               &~~~          &~10.5  &~26.9  &~0.16   &~      &~       ~&~1.567,p &~Q               &~   \\
 050210-205717*   &~0.47$\pm$0.13   &               &~~~          &    &       &       &~      &         & 1.025,p &~Q/Bz       &~  \\
 &~ & &~ &~ &~ &~ &~      &~ &~ &~FSRS     &~   \\
{\bf050555-293038}*&~0.83$\pm$0.24  &-0.79$\pm$0.33 &~{\bf1.3}  &~ 6.0  &~28.6  &~0.54   &~      &~5.4e28 ~&~3.750,s &~Q/Bz    &~  \\
&~ & &~ &~ &~ &~ &~      &~ &~ &~FSRS     &~   \\
{\bf050732-510416}&~1.44$\pm$0.19   &-0.29$\pm$0.02 &~~~{\bf1.2}  &~45.2  &~69.2  &~0.15   &~      &~1.3e26 ~&~0.529,s &~Q              &~t9,--SUM,PL\\
 051321-212821*   &~0.92$\pm$0.30   &               &~~~          &~  --  &~      &~      &~      &~       ~&~0.356,s &~G          &~  \\
 &~ & &~ &~ &~ &~ &~      &~ &~ &~FSRS     &~   \\
{\bf052350-441238}&~1.18$\pm$0.23   &-0.33          &~~~{\bf1.3}  &~ 8.2  &~14.3  &~0.16   &~      &~3.0e26 ~&~0.746,p &~Q        &~   \\
&~ & &~ &~ &~ &~ &~      &~ &~ &~FSRS?     &~   \\
{\bf053009-085132}&~1.26$\pm$0.26   &-0.23          &~~~{\bf1.3}  &~ 9.7  &~      &~0.11   &~      &~       ~&~        &~Q?            &~   \\
 054000-412746    &~0.72$\pm$0.11   &               &~~~          &~ 8.6  &~19.5  &~0.13   &~      &~       ~&~1.253,p &~Q              &~   \\
 054121-021108    &~0.86$\pm$0.07   &               &~~~          &~  --  &~      &~      &~      &~       ~&~        &~YSO             &~   \\
 054417-641914    &~0.93            &               &~~~          &~31.4  &~64.3  &~0.13   &~      &~       ~&~1.047,p &~Q               &~3f,\,t89 \\
{\bf054750-672801}&~1.11$\pm$0.12   &-0.43$\pm$0.04 &~~~{\bf~1.3  }&~8.4  &~16.7  &~0.10   &~      &~3.8e26 ~&~1.010,s &~Q/Bz            &~   \\
 054828-331331    &~0.87$\pm$0.11   &               &~~~          &~33.6  &~35.0  &~0.06   &~0.01  &~       ~&~0.041,s &~G               &~   \\
{\bf061540-011905}&~1.04$\pm$0.29   &-0.70          &~~~{\bf1.3}  &~ 5.6  &~      &~0.16   &~      &~       ~&~        &~G?              &~   \\
{\bf061826-533933}&~0.76            &-0.69$\pm$0.18 &~~~{\bf1.3}  &~ 4.7  &~~8.0  &~0.16   &~      &~3.1e26 ~&~0.695,p &~Q/Bz       &~   \\
&~ & &~ &~ &~ &~ &~      &~ &~ &~FSRS?     &~   \\
{\bf063004-551751}&~1.28            &-0.54          &~~~{\bf1.3}  &~ 6.7  &~~7.0  &~0.08   &~0.09  &~6.8e23 ~&~0.058,s &~G             &~   \\
 063345-135007    &~1.26$\pm$0.15   &               &~~~          &~32.9  &~      &~0.08   &~      &~       ~&~        &~G?             &~   \\
 063550-354658    &~0.86$\pm$0.05   &~              &~~~          &~12.9  &~35.7  &~0.09   &~      &~       ~&~1.769,p &~Q          &~   \\
 &~ & &~ &~ &~ &~ &~      &~ &~ &~FSRS     &~   \\
{\bf064929-592034}&~1.32$\pm$0.35   &-0.62$\pm$0.12 &~~~{\bf1.3}  &~ 7.3  &~24.0  &~0.16   &~      &~4.2e27 ~&~2.272,s &~Q/Bz      &~   \\
&~ & &~ &~ &~ &~ &~      &~ &~ &~FSRS?     &~   \\
 065057-651011    &~0.51            &-0.29          &~~~1.5       &~ 6.0  &~16.5  &~0.17   &~      &~1.8e27 ~&~1.756,p &~Q           &~   \\
 &~ & &~ &~ &~ &~ &~      &~ &~ &~FSRS     &~   \\
{\bf065532-171554}&~1.49$\pm$0.19   &-0.36          &~~~{\bf1.3}  &~ 8.0  &~      &~0.07   &~      &~       ~&~        &~G?             &~   \\
 065613-344658*   &~0.73$\pm$0.07   &               &~~~          &~43.8  &~      &~0.12   &~      &~       ~&~        &~G?             &~  \\
{\bf070903-785725}&~0.88            &-0.63          &~{\bf1.3}  &~ 4.8  &~      &~0.13   &~      &~       ~&~        &~G?         &~   \\
&~ & &~ &~ &~ &~ &~      &~ &~ &~FSRS?     &~   \\
 070949-381152    &~1.80            &               &~~~          &~20.6  &~23.2  &~0.09   &~0.02  &~       ~&~0.126,s &~G/Q            &~at,\,t89\\
               &~1.01            &               &~~~          &~14.5  &~16.3  &~0.09   &~      &~       ~&~      &~               &~   \\
{\bf071116-195103}&~1.11$\pm$0.08   &-0.44$\pm$0.07 &~{\bf1.3}  &~ 8.6  &~~8.7  &~0.17   &~      &~2.7e22 ~&~0.008,p &~PNe             &~   \\
 073630-041243    &~1.06$\pm$0.14   &               &~~~          &~71.5  &       &~0.15   &~      &~       ~&~        &~Q?              &~   \\
 073940-291118    &~1.10$\pm$0.12   &               &~~~          &~26.2  &~      &~0.12   &~      &~       ~&~        &~Q               &~t9 \\
 074109-544746    &~0.63$\pm$0.15   &               &~~~          &~   &~      &~      &~      &~       ~&~0.106,s &~G               &~   \\
 074554-004417*   &~0.72$\pm$0.08   &-0.63$\pm$0.09 &~~~1.5       &~ 7.7  &~15.3  &~1.55  &~      &~6.6e27 ~&~0.996,s &~Q/Bz    &~   \\
 &~ & &~ &~ &~ &~ &~      &~ &~ &~FSRS     &~   \\
{\bf080633-291135}&~1.19$\pm$0.13   &-0.77$\pm$0.01 &~~~{\bf1.3}  &~ 5.0  &~      &~0.27   &~      &~       ~&~        &~Q?      &~   \\
 080931-472011    &~0.80$\pm$0.11   &               &~~~          &~42.9  &~42.9  &~0.10   &~      &~       ~&~5.0E-5,s &~S               &~t89,\,PL\\
 081849-663400    &~0.81$\pm$0.18   &               &~~~          &~10.4  &~21.3  &~0.12   &~      &~       ~&~~1.055,p &~Q               &~   \\
{\bf082425-573631}&~0.83            &-0.41$\pm$0.10 &~~~{\bf1.2}  &~10.6  &~      &~0.07   &~      &~       ~&~         &~ ?             &~3f \\
 083046-170635    &~1.21$\pm$0.14   &               &~~~          &~49.7  &~      &~0.30   &~      &~       ~&~        &~Q/Bz?           &~t9   \\
 083529-595311    &~0.88$\pm$0.09   &-0.34$\pm$0.08 &~~~1.4       &~38.5  &~      &~0.51   &~      &~       ~&~        &~Q           &~t9,\,PL\\
 &~ & &~ &~ &~ &~ &~      &~ &~ &~FSRS     &~   \\
 083619-313331    &~0.93$\pm$0.15   &               &~~~          &~13.6  &~      &~0.12   &~      &~       ~&~        &~Q?             &~   \\
 083821-071336    &~0.61$\pm$0.08   &               &~~~          &~  --  &~      &~      &~      &~       ~&~0.117,p &~G              &~   \\
 084009-835432    &~0.76$\pm$0.18   &-0.53$\pm$0.16 &~~~1.4       &~ 4.8  &~~8.3  &~0.18   &~0.22  &~3.6e26 ~&~0.734,s &~G       &~   \\
 &~ & &~ &~ &~ &~ &~      &~ &~ &~FSRS     &~   \\
 084328-460641    &~0.65$\pm$0.09   &               &~~~          &~12.1  &~      &~0.11   &~      &~       ~&~        &~PNe            &~   \\
{\bf084511-652722}&~1.67$\pm$0.09   &-0.82$\pm$0.26 &~~~{\bf1.3}  &~ 7.9  &~26.0  &~0.43   &~      &~1.4e28 ~&~2.274,s &~Q       &~   \\
 084642-145620    &~0.92$\pm$0.20   &               &~~~          &~ 9.1  &~14.7  &~0.14   &~      &~       ~&~0.605,p &~Q               &~   \\
 085054-735144*   &~0.63$\pm$0.14   &               &~~~          &~   &~      &~      &~      &~       ~&~        &~Q?              &   \\
 085206-245534    &~2.03$\pm$0.08   &               &~~~          &~12.9  &~      &~0.10   &~      &~       ~&~        &~Q?              &~   \\
 090044-313128    &~0.71$\pm$0.11   &               &~~~          &~10.9  &~15.8  &~0.15   &~      &~       ~&~0.453,s &~Q           &~   \\
 &~ & &~ &~ &~ &~ &~      &~ &~ &~FSRS     &~   \\
 090420-311126*   &~0.66$\pm$0.05   &-0.59$\pm$0.08 &~~~1.5       &~52.5  &~      &~0.39   &~      &~       ~&~        &~Q/Bz?       &~  \\
 &~ & &~ &~ &~ &~ &~      &~ &~ &~FSRS     &~   \\
 090559-212012    &~1.14$\pm$0.03   &               &~~~          &~19.7  &~36.6  &~0.08   &~      &~       ~&~0.854,p &~Q               &~   \\
{\bf091900-253350}&~1.83$\pm$0.07   & -0.59          &~{\bf1.2}  &~ 7.0  &~~8.0  &~0.11   &~0.09  &~6.6e24 ~&~0.148,p &~G              &~   \\
 092051-872156    &~0.57            &-0.46$\pm$0.13 &~~~1.4       &~ 4.9  &~      &~0.08   &~      &~       ~&~       &~Q?              &~   \\
 093102-101325*   &~0.62$\pm$0.04   &-0.32$\pm$0.11 &~~~1.5       &~ 6.6  &~~9.9  &~0.12   &~      &~9.2e25 ~&~0.496,p &~Q/BL        &~TGSS,PL\\
 &~ & &~ &~ &~ &~ &~      &~ &~ &~FSRS     &~   \\
 093533-685722    &~0.96            &-0.19          &~~~1.4       &~ 8.5  &~14.7  &~0.09   &~      &~1.5e26 ~&~0.727,p &~Q              &~   \\
 093716-392518    &~0.74$\pm$0.11   &               &~~~          &~   &~      &~      &~      &~       ~&~        &~G?             &~   \\
 094219-231703*   &~0.49$\pm$0.05   &               &~~~          &~   &~      &~      &~      &~       ~&~1.547,s &~Q              &~  \\
 094258-604621    &~0.59            &               &~~~          &    &~      &~      &~      &~       ~&~0.586,s &~G               &~   \\
 095159-183703    &~1.21$\pm$0.09   &               &~~~          &~13.3  &~      &~0.08   &~      &~       ~&~        &~G?             &~t8 \\
 095440-594546    &~0.61$\pm$0.05   &               &~~~          &~   &~      &~      &~      &~       ~&~        &~G?              &~   \\
 095612-643928    &~0.50$\pm$0.04   &               &~~~          &~   &~      &~      &~      &~       ~&~0.803,p &~Q               &~   \\
 095633-404454*   &~0.99$\pm$0.01   &               &~~~          &~    &~      &~      &~      &~       ~&~1.414,s &~Q              &~t9 \\
 095727-015655    &~0.54            &-0.47$\pm$0.02 &~~~1.4       &~ 6.8  &~12.6  &~0.15   &~      &~1.4e26 ~&~0.860,s &~Q/Bz       &~    \\
 &~ & &~ &~ &~ &~ &~      &~ &~ &~FSRS     &~   \\
 095744-153246    &~0.65$\pm$0.09   &               &~~~          &~14.5  &~40.5  &~0.07   &~0.04  &~       ~&~1.787,s &~G           &~   \\
 &~ & &~ &~ &~ &~ &~      &~ &~ &~FSRS     &~   \\
 101112-221644    &~0.66$\pm$0.18   &               &~~~          &~   &~      &~      &~      &~       ~&~0.793,p &~Q              &~t9 \\
 101209-370129    &~0.46$\pm$0.03   & -0.04          &~~~1.6       &~12.1  &~      &~0.07   &~      &~       ~&~        &~G?            &~   \\
 101537-045440    &~0.52$\pm$0.03   &               &~~~          &~24.6  &~57.2  &~0.05   &~      &~       ~&~1.327,p &~Q?             &~   \\
 101956-002412    &~0.63$\pm$0.07   &               &~~~          &~30.0  &~63.7  &~0.06   &~      &~       ~&~1.125,s &~Q/Bz        &~   \\
 &~ & &~ &~ &~ &~ &~      &~ &~ &~FSRS     &~   \\
 102309-603240    &~0.60            & -0.20$\pm$0.02 &~~~1.5       &~ 7.1  &~      &~0.17   &~      &~       ~&~        &~PNe            &~   \\
 103003-192123    &~0.78$\pm$0.15   &               &~~~          &~10.5  &~15.7  &~0.06   &~      &~       ~&~0.500,p &~Q               &~   \\
 103504-173008*   &~0.91$\pm$0.19   &               &~~~          &~13.0  &~29.2  &~0.07   &~      &~       ~&~1.243,p &~Q/Bz?            &~TGSS\\
 103827-564707    &~0.57            & -0.07          &~~~1.5       &~ 8.0  &~      &~0.12   &~      &~       ~&~        &~PNe             &~   \\
 104227-210556    &~1.66            &               &~~~          &~   &~      &~      &~      &~       ~&~        &~?               &~t8 \\
 104416-535437    &~0.54            & -0.51          &~~~1.5       &~ 5.5  &~      &~0.08   &~      &~       ~&~        &~G              &~   \\
{\bf110019-651457}&~0.94$\pm$0.10   &               &~~~{\bf1.3}  &~ 9.2  &~~9.2  &~0.21   &~      &~       ~&~6.7E-5,s &~PNe            &    \\
 110317-512203    &~0.66$\pm$0.06   & -0.45          &~~~1.3       &~ 9.5  &~      &~0.10   &~      &~       ~&~       &~Q?             &~3f \\
 110436-540055    &~0.93            & -0.47          &~~~1.4       &~ 5.6  &~      &~0.08   &~      &~       ~&~        &~Q              &~   \\
{\bf110828-123121}&~1.17$\pm$0.11   & -0.18          &~~~{\bf1.3}  &~ 8.0  &~29.6  &~0.06   &~      &~1.3e27 ~&~2.687,p &~Q/BL           &~   \\
 110918-481518*   &~0.50$\pm$0.03   & -0.46$\pm$0.04 &~~~1.5       &~44.0  &173.2  &~0.36   &~      &~1.3e28 ~&~2.937,p &~Q/Bz?      &~PL \\
 &~ & &~ &~ &~ &~ &~      &~ &~ &~FSRS     &~   \\
 110957-373220*   &~0.84            &               &~~~          &~48.6  &~49.1  &~0.08   &~0.01  &~       ~&~0.010,s &~G           &~  \\
 &~ & &~ &~ &~ &~ &~      &~ &~ &~FSRS     &~   \\
 111015-665531    &~1.28$\pm$0.06   &               &~~~          &~23.8  &~      &~0.14   &~      &~       ~&~        &~Q              &~3f,t89\\
 111228-625857    &~0.73$\pm$0.01   &               &~~~          &~26.8  &~      &~0.16   &~      &~       ~&~        &~G              &~   \\
 111246-203932    &~1.80$\pm$0.05   &               &~~~          &~48.8  &~      &~0.12   &~      &~       ~&~        &~G?             &~t89\\
 111430-860325    &~0.53$\pm$0.12   & -0.46$\pm$0.13 &~~~1.5       &~ 5.6  &~      &~0.15   &~      &~       ~&~        &~Q?          &~   \\
 &~ & &~ &~ &~ &~ &~      &~ &~ &~FSRS     &~   \\
 111605-263758    &~1.75$\pm$0.10   &               &~~~          &~   &~      &~      &~      &~       ~&~        &~G?             &~t89\\
 111719-483809*   &~0.49$\pm$0.07   & -0.25$\pm$0.07 &~~~1.4       &~29.2  &~      &~0.17   &~      &~       ~&~        &~Q/Bz           &   \\
{\bf112120-172242}&~1.12$\pm$0.16   & -0.90$\pm$0.10 &~~~{\bf1.3}  &~ 6.1  &~12.1  &~0.21   &~      &~1.1e27 ~&~0.986,s &~Q/BL           &~--WISH\\
                &~0.84$\pm$0.05   &              &~~~1.5       &~ 5.8  &~11.5  &~0.18   &~      &~       ~&~           &~         &~+WISH\\
 112621-312358    &~0.94$\pm$0.17   &               &~~~          &~11.0  &~14.5  &~0.07   &~0.05  &~       ~&~0.321,s &~G     ~         &~   \\
{\bf112931-443552}&~1.01$\pm$0.11   & -0.73$\pm$0.10 &~~~{\bf1.3}  &~ 4.8  &~~6.3  &~0.20   &~      &~6.5e25 ~&~0.317,s &~Q/Bz?   &~   \\
&~ & &~ &~ &~ &~ &~      &~ &~ &~FSRS?     &~   \\
{\bf112953-024006}&~0.94$\pm$0.15   & -0.67          &~~~{\bf1.3}  &~ 9.2  &~28.4  &~0.19   &~      &~4.4e27 ~&~2.087,s &~Q         &~    \\
&~ & &~ &~ &~ &~ &~      &~ &~ &~FSRS?     &~   \\
 113143-581853    &~0.73$\pm$0.13   & -0.73$\pm$0.03 &~~~1.5       &~ 9.2  &~      &~1.30  &~      &~       ~&~        &~Q      &~PL \\
 113316-631726    &~0.61$\pm$0.14   & -0.28$\pm$0.11 &~~~1.5       &~ 5.7  &~      &~0.21   &~      &~       ~&~         &~G?              &~   \\
{\bf113724-822905}&~1.17$\pm$0.05   & -0.59          &~~~{\bf1.3}  &~ 9.1  &~      &~0.09   &~      &~       ~&~        &~Q?              &~   \\
 114002-464103    &~0.55$\pm$0.01   & -0.35          &~~~1.5       &~ 8.5  &~26.1  &~0.08   &~      &~1.3e27 ~&~2.055,s &~Q              &~   \\
 114011-680649    &~0.70$\pm$0.10   &               &~~~          &~12.1  &~      &~0.10   &~      &~       ~&~        &~G?             &~   \\
 114503-325824    &~0.61$\pm$0.09   & -0.22          &~~~1.5       &~13.6  &~14.1  &~0.08   &~0.03  &~2.8e23 ~&~0.038,s &~G              &~   \\
 114838-650837    &~0.53            & -0.08          &~~~1.5       &~ 6.5  &~      &~0.07   &~      &~       ~&~        &~PNe             &~   \\
 114844-781933    &~1.19$\pm$0.01   &               &~~~          &~23.9  &~      &~0.13   &~      &~       ~&~        &~Q               &~3f,t89\\
 115031-842623    &~0.68            &               &~~~          &~   &~      &~      &~      &~       ~&~0.309,p &~Q               &~at,--SUM\\
{\bf115034-541642}&~1.60$\pm$0.05   & -0.57$\pm$0.11 &~~~{\bf1.3}  &~ 7.6  &~      &~0.32   &~      &~       ~&~        &~Q      &~   \\
 115503-310759    &~0.76$\pm$0.06   & -0.52$\pm$0.06 &~~~1.4       &~ 6.1  &~32.4  &~0.18   &~      &~1.6e28 ~&~4.308,s &~Q/Bz        &~   \\
 &~ & &~ &~ &~ &~ &~      &~ &~ &~FSRS     &~   \\
 115546-761907    &~0.63            & -0.54$\pm$0.09 &~~~1.4       &~ 3.9  &~      &~0.12   &~      &~       ~&~        &~Q?             &~   \\
 115918-663539    &~0.70$\pm$0.32   & -0.58$\pm$0.13 &~~~1.4       &~ 4.1  &~      &~0.31   &~      &~       ~&~        &~Q     ~&~   \\
{\bf115951-214853}&~1.19$\pm$0.20   & -0.77$\pm$0.14 &~~~{\bf1.3}  &~ 5.1  &~~9.8  &~0.81   &~      &~3.2e27 ~&~0.927,s &~Q       &~   \\
&~ & &~ &~ &~ &~ &~      &~ &~ &~FSRS?     &~   \\
 120458-505556    &~0.72$\pm$0.10   &               &~~~          &~16.3  &~33.9  &~0.23   &~      &~       ~&~1.083,s &~Q/Bz?       &~   \\
 &~ & &~ &~ &~ &~ &~      &~ &~ &~FSRQ     &~   \\
 121255-175345*   &~1.13$\pm$0.13   &               &~~~          &~14.6  &~37.5  &~0.11   &~      &~       ~&~1.575,s &~Q               &~  \\
 121340-272423*   &~0.97$\pm$0.13   &               &~~~          &~12.4  &~      &~0.16   &~      &~       ~&~        &~Q/Bz?           &   \\
 122033-273601*   &~0.54$\pm$0.04   &               &~~~          &~     &~      &~      &~      &~       ~&~0.912,p &~Q           &~TGSS\\
 &~ & &~ &~ &~ &~ &~      &~ &~ &~FSRS     &~   \\
 122609-294012*   &~0.81            & -0.31          &~~~          &~10.6  &~      &~0.08   &~      &~       ~&~        &~Q/Bz       &~at,TGSS\\
 &~ & &~ &~ &~ &~ &~      &~ &~ &~FSRQ?     &~   \\
 122635-190438*   &~0.65$\pm$0.05   & -0.27          &~~~1.4       &~12.0  &~20.6  &~0.31   &~      &~5.2e26 ~&~0.719,p &~Q              &~  \\
 123030-645206    &~0.83$\pm$0.13   &               &~~~          &~ 9.3  &~      &~0.14   &~      &~       ~&~        &~PNe            &~   \\
 123449-243232    &~0.66$\pm$0.03   & -0.42          &~~~1.4       &~10.5  &~12.4  &~0.09   &~0.05  &~8.1e24 ~&~0.181,s &~G/BL           &~   \\
 124114-273026*   &~0.60$\pm$0.07   &               &~~~          &~29.6  &101.3  &~0.08   &~      &~       ~&~2.427,p &~Q              &~  \\
 125437-200056*   &~0.50$\pm$0.08   & -0.25$\pm$0.15 &~~~1.4       &~25.3  &~49.6  &~0.32   &~      &~9.7e26 ~&~0.959,s &~Q              &~  \\
 125748-254802    &~0.72$\pm$0.13   &               &~~~          &~11.1  &~36.3  &~0.09   &~      &~       ~&~2.273,p &~Q              &~   \\
 130031-441442*   &~0.77$\pm$0.09   & -0.34$\pm$0.17 &~~~1.4       &~ 7.1  &~~7.3  &~0.11   &~0.09  &~2.7e23 ~&~0.032,s &~G              &~TGSS\\
 130340-462103*   &~0.74$\pm$0.14   & -0.47$\pm$0.11 &~~~1.4       &~ 5.5  &~14.6  &~0.22   &~      &~2.5e27 ~&~1.664,s &~Q/Bz   &~   \\
 &~ & &~ &~ &~ &~ &~      &~ &~ &~FSRQ     &~   \\
                &~0.57$\pm$0.14   & -0.47$\pm$0.11 &~~~1.8       &~ 8.3  &~22.1  &~0.17   &~      &~2.0e27 ~&~       &~             &~+TGSS\\
 130713-431426*   &~0.75            & -0.86          &~~~1.4       &~ 6.1  &~20.3  &~0.09   &       &~3.4e27 ~&~2.343,p &~Q               &~TGSS\\
 134229-740728    &~0.51$\pm$0.12   &               &~~~          &~   &~      &~      &~      &~       ~&~        &~G?            &~--SUM\\
{\bf140003-185811}*&~1.11$\pm$0.05  & -0.61$\pm$0.09 &~~~{\bf1.3}  &~ 6.9  &~13.4  &~0.41   &~      &~1.5e27 ~&~0.940,s &~Q/BL       &~ \\
&~ & &~ &~ &~ &~ &~      &~ &~ &~FSRQ     &~   \\
 140257-664031    &~1.66$\pm$0.29   &               &~~~          &~13.3  &~      &~0.09   &~      &~       ~&~        &~Q?             &~3f,t89\\
 141601-292450    &~0.65            &               &~~~          &~10.2  &~      &~0.14   &~      &~       ~&~        &~Q?             &~   \\
{\bf141912-262730}* &~0.74$\pm$0.18  & -0.32          &~~~{\bf1.3}  &~11.8  &~27.7  &~0.11   &~      &~7.2e26 ~&~1.358,p &~Q          &~  \\
&~ & &~ &~ &~ &~ &~      &~ &~ &~FSRS?     &~   \\
 141922-083830*   &~0.63$\pm$0.10   & -0.52$\pm$0.06 &~~~1.4       &~23.1  &~43.8  &~0.36   &~      &~1.1e27 ~&~0.903,s &~Q/Bz      &~PL \\
 &~ & &~ &~ &~ &~ &~      &~ &~ &~FSRQ    &~   \\
 142119-583822    &~0.73$\pm$0.06   &               &~~~          &~13.0  &~      &~0.07   &~      &~       ~&~        &~PNe            &~   \\
 142741-330531*   &~0.63$\pm$0.05   & -0.40$\pm$0.03 &~~~1.4       &~89.8  &156.3  &~1.63  &~      &~3.2e27 ~&~0.742,s &~Q/Bz       &~  \\
 &~ & &~ &~ &~ &~ &~      &~ &~ &~FSRS     &~   \\
 143608-153609    &~1.06$\pm$0.07   &               &~~~          &~   &~      &~      &~      &~       ~&~        &~Q?             &~t9\\
 144458-085941*   &~0.55$\pm$0.01   &               &~~~          &~    &~      &~      &~      &~       ~&~1.788,p &~Q/Bz           &~TGSS\\
 144555-303705*   &~0.69$\pm$0.09   &               &~~~          &~    &~      &~      &~      &~       ~&~0.573,p &~Q          &~t9\\
 &~ & &~ &~ &~ &~ &~      &~ &~ &~FSRS     &~   \\
 145508-315832*   &~0.48$\pm$0.06   &               &~~~          &~   &~      &~      &~      &~       ~&~        &~Q/Bz?       &~  \\
 &~ & &~ &~ &~ &~ &~      &~ &~ &~FSRS     &~   \\
 150930-264734*   &~0.89$\pm$0.13   &               &~~~          &~15.9  &~79.2  &~0.13   &~      &~       ~&~3.974,p &~?          &~  \\
 &~ & &~ &~ &~ &~ &~      &~ &~ &~FSRS?     &~   \\
 151330-422156*   &~0.78$\pm$0.15   & -0.48          &~~~ 1.4      &~ 7.9  &~      &~0.30   &~      &~       ~&~        &~Q/Bz            &~  \\
 152210-294830*   &~0.86$\pm$0.13   &               &~~~          &~12.5  &~19.1  &~0.18   &~      &~       ~&~0.532,s &~Q/BL           &~  \\
                &~1.30$\pm$0.31   &               &~~~          &~12.2  &~18.6  &~0.19   &~      &~       ~&~             &~         &~  \\
 152445-510353    &~0.58            & -0.30          &~~~1.5       &~12.9  &~      &~0.07   &~      &~       ~&~        &~Q?             &~   \\
 153030-220811    &~0.94$\pm$0.05   &               &~~~          &~10.5  &~      &~0.07   &~      &~       ~&~        &~Q/Bz           &~t9\\
            &~1.52$\pm$0.07   &               &~~~          &~24.1  &~      &~0.08   &~      &~       ~&~        &~              &~at \\
 153744-295433    &~1.48$\pm$0.21   &               &~~~          &~20.0  &~      &~0.13   &~      &~       ~&~        &~G?             &~t9 \\
{\bf153851-165526}&~1.27$\pm$0.16   & -0.49          &~~~{\bf1.3}  &~ 8.9  &~      &~0.11   &~      &~       ~&~        &~Q/Bz           &  \\
 154120-181521*   &~0.95$\pm$0.22   &               &~~~          &~15.1  &~26.7  &~0.08   &~      &~       ~&~0.771,p &~Q/Bz           &~TGSS\\
 155355-235841    &~0.90$\pm$0.12   &               &~~~          &~20.9  &~      &~0.11   &~      &~       ~&~        &~S              &~    \\
 155704-272446*   &~0.82$\pm$0.08   &               &~~~          &~10.8  &~19.9  &~0.19   &~      &~       ~&~0.836,s &~Q          &~  \\
 &~ & &~ &~ &~ &~ &~      &~ &~ &~FSRS?     &~   \\
 155941-244240    &~0.70$\pm$0.09   & -0.49$\pm$0.02 &~~~1.6       &~ 1.8  &~ 3.2  &~0.58   &~      &~2.0e28 ~&~2.820,s &~Q/BL       &~  \\
 &~ & &~ &~ &~ &~ &~      &~ &~ &~FSRQ     &~   \\
 155954-175859    &~1.06$\pm$0.11   &               &~~~          &~ 9.9  &~      &~0.10   &~      &~       ~&~        &~Q?         &~   \\
 &~ & &~ &~ &~ &~ &~      &~ &~ &~FSRS?     &~   \\
 161845-142428    &~1.42$\pm$0.17   &               &~~~          &~32.4  &~39.1  &~0.09   &~      &~       ~&~0.206,p &~Q?             &~3f,t8\\
 162125-383707    &~1.42$\pm$0.03   &               &~~~          &~13.7  &~      &~0.26   &~      &~       ~&~        &~Q/Bz?          &~ \\
 163741-381212    &~0.51$\pm$0.08   &               &~~~          &~    &~      &~      &~      &~       ~&~        &~G?             &~   \\
 163827-170111    &~1.24$\pm$0.33   &               &~~~          &~ 8.3  &~      &~0.22   &~      &~       ~&~       &~Q?         &~   \\
 &~ & &~ &~ &~ &~ &~      &~ &~ &~FSRS?     &~   \\
 164516-331816*   &~0.53$\pm$0.06   &               &~~~          &~33.3  &~52.5  &~0.35   &~      &~       ~&~0.576,p &~Q/Bz?           &~  \\
 164842-330147*   &~0.58$\pm$0.06   & -0.36$\pm$0.02 &~~~1.5       &~19.1  &~      &~0.86   &~      &~       ~&~        &~Q               &~  \\
 164854-354707    &~0.84$\pm$0.12   &               &~~~          &~26.8  &~26.8  &~0.12   &~      &~       ~&~3.4E-4,s &~PNe             &~   \\
 170537-283810*   &~0.50$\pm$0.05   &               &~~~          &~   &~      &~      &~      &~       ~&~        &~Q?              &~TGSS\\
           &~0.93$\pm$0.01   &               &~~~1.4       &~18.3  &~      &~0.22   &~      &~       ~&~       &~              &~   \\
 171043-471820    &~1.06$\pm$0.06   &               &~~~          &~35.9  &~      &~0.08   &~      &~       ~&~        &~?              &~t8 \\
 172654-513801*   &~0.50$\pm$0.01   &               &~~~          &~    &~      &~      &~      &~       ~&~        &~Q?             &~TGSS\\
 172746-754617*   &~0.76$\pm$0.02   &               &~~~          &~23.9  &~53.3  &~0.07   &~      &~       ~&~1.227,p &~Q              &~t9,--SUM\\
 174220-343515    &~0.61$\pm$0.04   &               &~~~          &~    &~      &~      &~      &~       ~&~        &~G?             &~   \\
 180754-641350*   &~0.57$\pm$0.08   & -0.39$\pm$0.04 &~~~1.6       &~21.9  &~44.2  &~0.38   &~      &~1.4e27 ~&~1.016,s &~Q         &~  \\
 &~ & &~ &~ &~ &~ &~      &~ &~ &~FSRS     &~   \\
{\bf181225-712006}&~0.69            & -0.54$\pm$0.23 &~~~{\bf1.3}  &~13.0  &~15.6  &~0.04   &~0.03  &~4.5e24 ~&~0.199,s &~G              &~   \\
 181436-641253*   &~0.58$\pm$0.16   &               &~~~          &~33.3  &~51.2  &~0.13   &~      &~       ~&~0.539,p &~Q              &   \\
{\bf181612-305208}&~1.05$\pm$0.08   & -0.57$\pm$0.09 &~~~{\bf1.3}  &~15.9  &~15.9  &~0.21   &~      &~3.6e17 ~&~2.7E-5,s &~PNe             &~   \\
 182016-424342*   &~0.61$\pm$0.13   &               &~~~          &~   &~      &~      &~      &~       ~&~2.896,p &~Q?        & --SUM\\
 &~ & &~ &~ &~ &~ &~      &~ &~ &~FSRS     &~   \\
 183923-345348    &~1.73$\pm$0.16   &               &~~~          &~30.3  &~      &~0.32   &~      &~       ~&~        &~Q?              &~t9 \\
 184114-450124    &~0.61$\pm$0.18   &               &~~~          &~16.1  &~      &~0.08   &~      &~       ~&~        &~G?             &~   \\
{\bf184347-793646}&~1.07$\pm$0.15   & -0.76          &~~~{\bf1.3}  &~ 6.5  &~21.6  &~0.12   &~      &~4.1e27 ~&~2.352,p &~Q              &~   \\
{\bf184723-162302}&~1.21$\pm$0.14   &-0.63          &~~~{\bf1.3}  &~ 9.2  &~      &~0.09   &~      &~       ~&~        &~Q?             &~   \\
 184827-735337    &~0.64$\pm$0.04   &               &~~~          &~    &~      &~      &~      &~       ~&~        &~G          &~\\
 &~ & &~ &~ &~ &~ &~      &~ &~ &~FSRS     &~   \\
 185910-045825    &~0.86$\pm$0.18   &               &~~~          &~   &~      &~      &~      &~       ~&~        &~G?             &~\\
 190510-093442    &~0.51$\pm$0.02   &               &~~~          &~   &~      &~      &~      &~       ~&~        &~Q?             &~\\
 191706-600015*   &~0.52$\pm$0.004  &               &~~~          &~   &~      &~      &~      &~       ~&~        &~Q?             &~  \\
 191816-411131*   &~0.76$\pm$0.03   &               &~~~          &~   &~      &~      &~      &~       ~&~1.217,p &~Q/BL        &~t9\\
 &~ & &~ &~ &~ &~ &~      &~ &~ &~FSRS     &~   \\
 191843-293110    &~1.26$\pm$0.02   &               &~~~          &~12.4  &~      &~0.06   &~      &~       ~&~        &~G?            &~   \\
 191931-180635*   &~0.98$\pm$0.04   &               &~~~          &~18.7  &~31.5  &~0.15   &~      &~       ~&~0.689,p &~Q          &   \\
 &~ & &~ &~ &~ &~ &~      &~ &~ &~FSRS     &~   \\
 192809-203544    &~0.90$\pm$0.04   &               &~~~          &~12.2  &~21.4  &~0.31   &~      &~       ~&~0.747,p &~Q/Bz       &~    \\
 &~ & &~ &~ &~ &~ &~      &~ &~ &~FSRS     &~   \\
 194131-760548    &~0.71$\pm$0.02   &               &~~~          &~  &~      &~      &~      &~       ~&~0.146,s &~G/BL           &~    \\
 195739-461113    &~0.54$\pm$0.05   &               &~~~          &~    &~      &~      &~      &~       ~&~       &~Q?        &~   \\
 &~ & &~ &~ &~ &~ &~      &~ &~ &~FSRS     &~   \\
 195949-441611    &~0.62$\pm$0.03   &               &~~~          &~   &~      &~      &~      &~       ~&~1.667,p &~Q              &~t9 \\
 200012-474951    &~0.75$\pm$0.01   &               &~~~          &~20.6  &~43.4  &~0.08   &~      &~       ~&~1.109,p &~Q              &~   \\
 200324-042137    &~0.72$\pm$0.33   &-0.43$\pm$0.01 &~~~1.4       &~ 6.1  &~      &~0.15   &~      &~       ~&~        &~Q      &~   \\
 &~ & &~ &~ &~ &~ &~      &~ &~ &~FSRS     &~   \\
 201115-154640*   &~0.53$\pm$0.01   & -0.52$\pm$0.03 &~~~1.5       &~20.8  &~45.3  &~1.89  &~      &~1.1e28 ~&~1.180,s &~Q      &~  \\
 &~ & &~ &~ &~ &~ &~      &~ &~ &~FSRS     &~   \\
 201500-671258    &~0.76$\pm$0.11   &               &~~~          &~15.3  &~18.9  &~0.22   &~      &~       ~&~0.239,s &~Q/BL        &~   \\
 &~ & &~ &~ &~ &~ &~      &~ &~ &~FSRS     &~   \\
{\bf203540-694407}&~1.20$\pm$0.13   & -0.62          &~~~{\bf1.3}  &~ 8.7  &~16.3  &~0.23   &~      &~6.3e26 ~&~0.877,p &~Q              &~   \\
 203637-283027*   &~0.77$\pm$0.23   &               &~~~          &~14.8  &~48.8  &~0.30   &~      &~       ~&~2.308,s &~Q/Bz           &~  \\
 204849-254615    &~1.09$\pm$0.11   &               &~~~          &~19.7  &~37.5  &~0.17   &~      &~       ~&~0.900,p &~Q              &~   \\
{\bf205503-635207}&~1.35$\pm$0.21   & -0.63          &~~~{\bf1.2}  &~18.7  &~44.5  &~0.04   &~      &~3.5e26 ~&~1.377,p &~Q              &~   \\
 205625-320845*   &~0.68$\pm$0.03   & -0.53$\pm$0.11 &~~~1.5       &~39.1  &145.5  &~0.77   &~      &~2.7e28 ~&~2.727,p &~Q/Bz           &   \\
 210923-261409    &~0.51$\pm$0.10   &               &~~~          &~23.6  &~      &~0.14   &~      &~       ~&~        &~G?             &~   \\
 210925-361557    &~0.92$\pm$0.08   &               &~~~          &~14.6  &~17.9  &~0.07   &~0.03  &~       ~&~0.226,s &~G               &~   \\
 212402-602808*   &~0.55            & -0.36$\pm$0.14 &~~~1.6       &~12.5  &~26.2  &~0.10   &~      &~4.3e26 ~&~1.098,s &~Q/Bz           &~  \\
{\bf212642-381832}*&~0.77$\pm$0.30  & -0.37          &~~~{\bf1.3}  &~ 9.5  &~      &~0.10   &~      &~       ~&~      &~Q?              &~  \\
 213208-542037*   &~0.57$\pm$0.07   & -0.60$\pm$0.13 &~~~1.4       &~36.5  &~69.7  &~0.20   &~      &~6.8e26 ~&~0.909,p &~Q              &~PL\\
 213314-111517*   &~1.14$\pm$0.31   &               &~~~          &~ 9.8  &~15.3  &~0.06   &~      &~       ~&~0.564,p &~Q              &~  \\
 214447-694654    &~0.51$\pm$0.01   & -0.37          &~~~1.6       &~33.2  &182.9  &~0.13   &~      &~9.4e27 ~&~4.517,p &~Q              &~   \\
 215624-173401    &~0.64$\pm$0.06   &               &~~~          &~13.2  &~29.0  &~0.07   &~      &~       ~&~1.200,p &~Q?              &~   \\
 220016-371656*   &~0.49$\pm$0.05   &               &~~~          &~31.5  &110.9  & 0.23   &~      &~       ~&~2.518,p &~Q          &   \\
 &~ & &~ &~ &~ &~ &~      &~ &~ &~FSRS     &~   \\
 220413-465424    &~1.96            &               &~~~          &~13.3  &~24.3  &~0.12   &~0.05  &~       ~&~0.832,p &~G?             &~3f \\
 221418-400849    &~0.84$\pm$0.04   &               &~~~          &~17.9  &~23.2  &~0.19   &~      &~       ~&~0.293,p &~Q/Bz           &~   \\
 222329-061201*   &~0.63$\pm$0.09   &               &~~~          &~   &~      &~      &~      &~       ~&~1.349,p &~Q?              &   \\
 223015-132543*   &~0.50$\pm$0.04   & -0.65$\pm$0.08 &~~~1.4       &~14.1  &~34.0  &~0.75   &~      &~7.2e27 ~&~1.420,s &~Q/Bz        &~PL\\
 &~ & &~ &~ &~ &~ &~      &~ &~ &~FSRS     &~   \\
 224500-493148*   &~0.71            &               &~~~          &~31.8  &~63.6  &~0.10   &~      &~       ~&~1.003,s &~Q              &~3f\\
 224752-123719*   &~0.69$\pm$0.10   & -0.66$\pm$0.09 &~~~1.5       &~15.3  &~44.2  &~0.44   &~      &~8.2e27 ~&~1.892,s &~Q/Bz        &~PL\\
 &~ & &~ &~ &~ &~ &~      &~ &~ &~FSRQ     &~   \\
 225423-514844*   &~0.66            &               &~~~          &~15.6  &~20.6  &~0.06   &~      &~       ~&~0.316,p &~Q/BL           &~at\\
 225558-103922    &~0.80$\pm$0.10   &               &~~~          &~14.1  &~22.8  &~0.10   &~0.04  &~       ~&~0.616,p &~G?             &~   \\
 230737-354828    &~0.62$\pm$0.02   & -0.40          &~~~1.5       &~19.8  &~23.9  &~0.17   &~      &~9.4e26 ~&~1.207,p &~Q              &    \\
 231339-183004    &~1.02$\pm$0.03   &               &~~~          &~16.3  &~      &~0.08   &~      &~       ~&~        &~G?             &~   \\
 231347-441615*   &~0.76$\pm$0.01   &               &~~~          &~61.6  &118.2  &~0.18   &~      &~       ~&~0.919,p &~Q              &~t9,3f\\
 231546-230744    &~0.93$\pm$0.17   &               &~~~          &~12.5  &~13.3  &~0.08   &~0.04  &~       ~&~0.069,s &~G              &~   \\
 233159-381147    &~0.86$\pm$0.13   & -0.78$\pm$0.12 &~~~1.6       &~ 3.9  &~~8.6  &~0.67   &~      &~4.9e27 ~&~1.195,p &~Q              &~   \\
 233726-590113    &~0.78$\pm$0.13   &               &~~~          &~34.6  &~57.8  &~0.28   &~      &~       ~&~0.672,p &~Q/Bz?          &~   \\
 234540-720354*   &~0.74            &               &~~~{\bf1.3}  &~11.7  &~22.9  &~0.05   &~      &~       ~&~0.958,p &~Q/Bz?          &~at \\
 234743-494627*   &~0.63$\pm$0.16   & -0.32$\pm$0.10 &~~~1.6       &~23.1  &~37.9  &~0.38   &~      &~5.0e26 ~&~0.643,s &~Q           &~PL \\
 &~ & &~ &~ &~ &~ &~      &~ &~ &~FSRQ    &~   \\
 235311-274324*   &~1.02$\pm$0.10   &               &~~~          &~16.6  &~31.4  &~0.28   &~      &~       ~&~0.889,s &~Q/Bz            &  \\
 235540-541834    &~0.64$\pm$0.09   &               &~~~          &~30.6  &~43.3  &~0.10   &~      &~       ~&~0.413,s &~Q              &~ \\

\end{longtable*}
\end{enumerate}

\subsection{Spectra} \label{Fig}
Examples of the most characteristic spectra of some HFP radio sources.
\begin{figure*}
\centerline{
\includegraphics[angle=0,width=1.0\textwidth,clip]{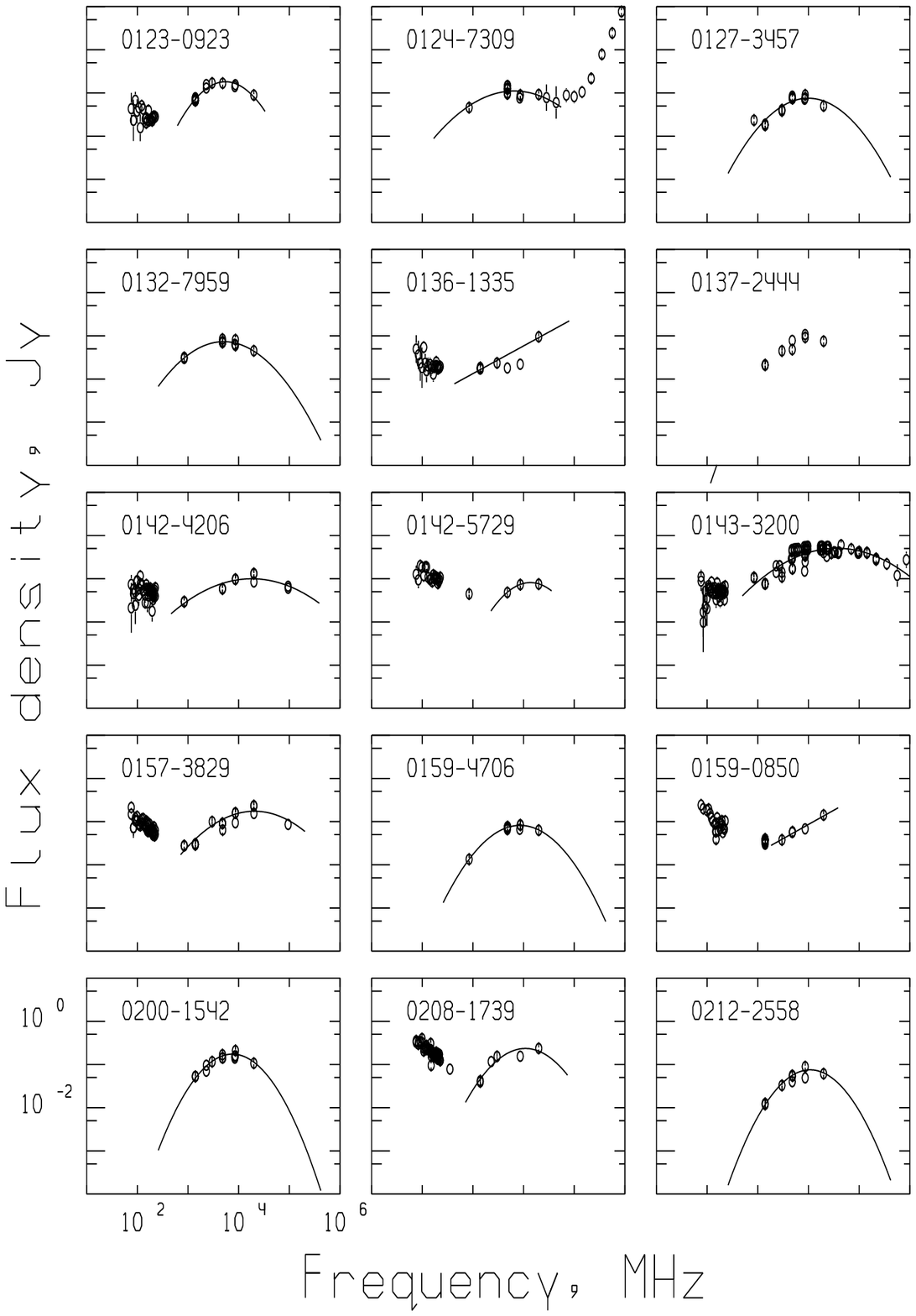}
}
\caption{
Examples of spectra of sources of the formed sample.
}
\label{fig17}
\end{figure*}
\begin{figure*}
\centerline{
\includegraphics[angle=0,width=1.0\textwidth,clip]{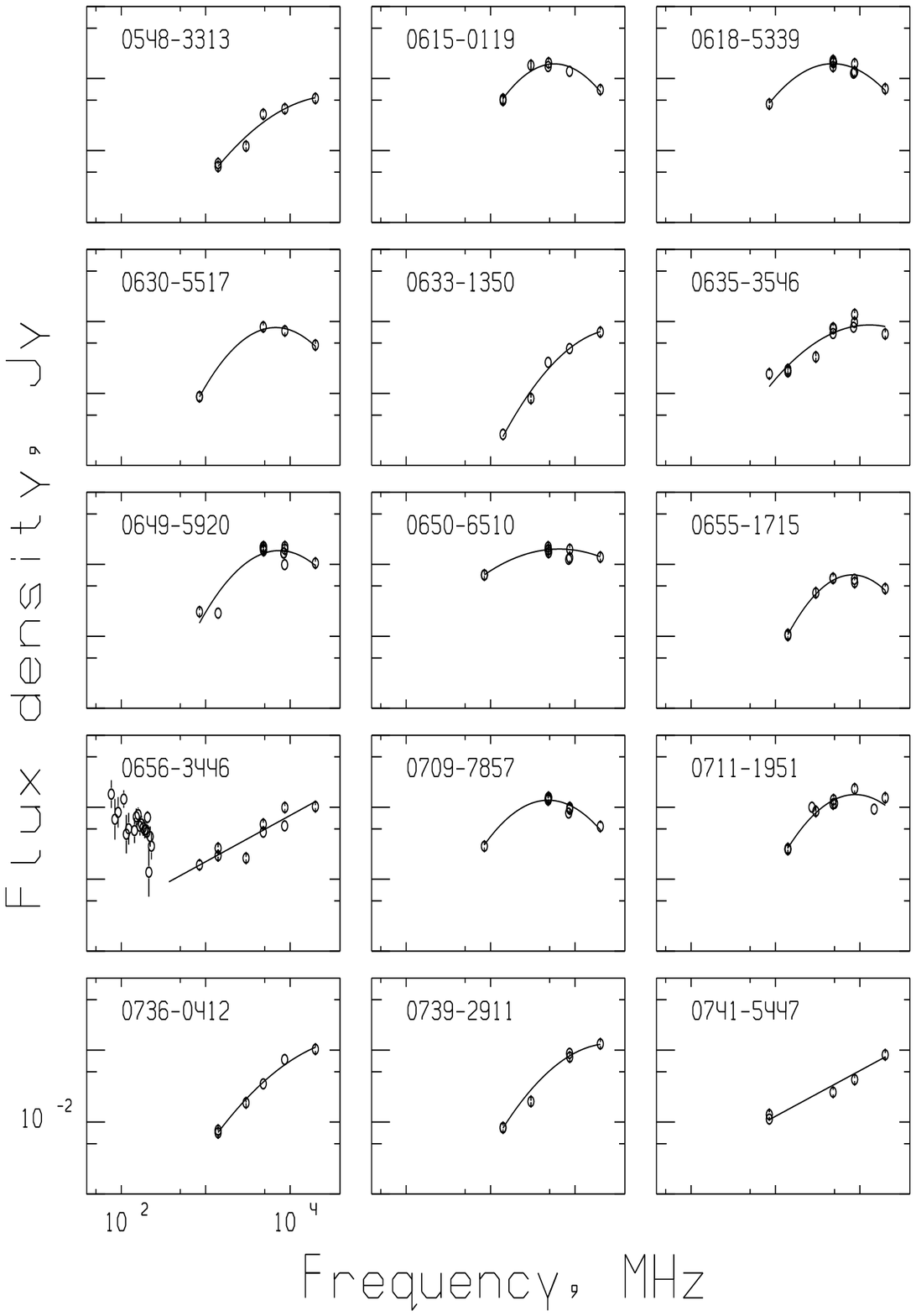}
}
\caption{
Examples of spectra of sources of the formed sample.
}
\label{fig18}
\end{figure*}
\begin{figure*}
\centerline{
\includegraphics[angle=0,width=1.0\textwidth,clip]{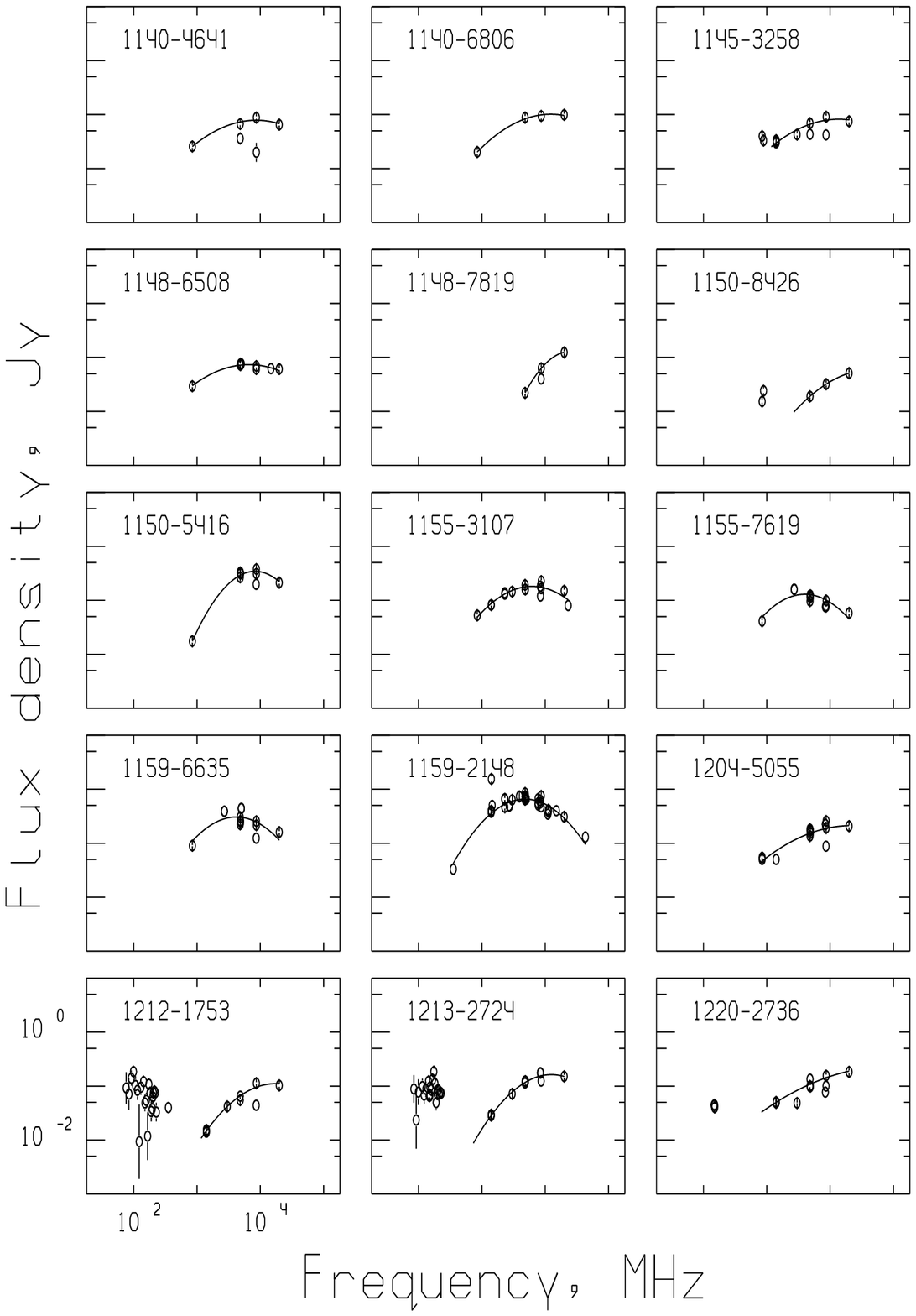}
}
\caption{
Examples of spectra of sources of the formed sample.
}
\label{fig19}
\end{figure*}

\begin{figure*}
\centerline{
\includegraphics[angle=0,width=1.0\textwidth,clip]{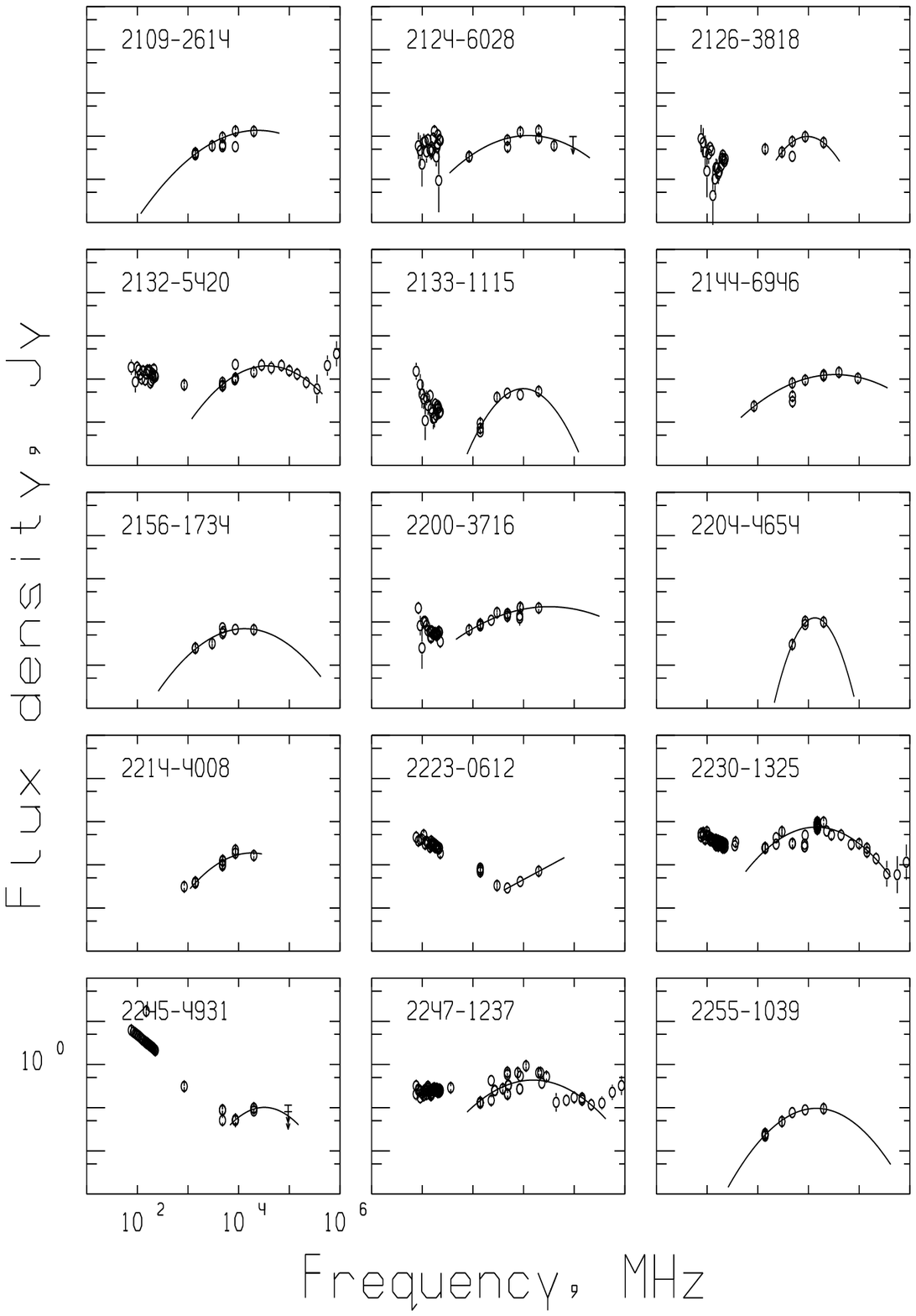}
}
\caption{
Examples of spectra of sources of the formed sample.
}
\label{fig20}
\end{figure*}

\end{document}

%% file: sao_cmd_author.tex
\def\saoname{Special Astrophysical Observatory,  Russian Academy of Sciences,
              Nizhnii Arkhyz, 369167 Russia}

\def\squareforqed{\hbox{\rlap{$\sqcap$}$\sqcup$}}

\def\sq{\ifmmode\squareforqed\else{\unskip\nobreak\hfil
\penalty50\hskip1em\null\nobreak\hfil\squareforqed
\parfillskip=0pt\finalhyphendemerits=0\endgraf}\fi}

\def\utw{\smash{\rlap{\lower5pt\hbox{$\sim$}}}}

\def\udtw{\smash{\rlap{\lower6pt\hbox{$\approx$}}}}

\def\diameter{{\ifmmode\mathchoice
{\ooalign{\hfil\hbox{$\displaystyle/$}\hfil\crcr
{\hbox{$\displaystyle\mathchar"20D$}}}}
{\ooalign{\hfil\hbox{$\textstyle/$}\hfil\crcr
{\hbox{$\textstyle\mathchar"20D$}}}}
{\ooalign{\hfil\hbox{$\scriptstyle/$}\hfil\crcr
{\hbox{$\scriptstyle\mathchar"20D$}}}}
{\ooalign{\hfil\hbox{$\scriptscriptstyle/$}\hfil\crcr
{\hbox{$\scriptscriptstyle\mathchar"20D$}}}}
\else{\ooalign{\hfil/\hfil\crcr\mathhexbox20D}}%
\fi}}